\documentclass{jfm}
\usepackage{amsmath}
\usepackage{verbatim}
\usepackage{bm}
\usepackage{xcolor}
\usepackage[caption=false]{subfig}
\usepackage{rotating}
\usepackage{booktabs}
\usepackage{multirow}
\usepackage[normalem]{ulem}

\definecolor{darkgreen}{rgb}{0.0,0.5,0.0}
\usepackage[colorlinks,citecolor=darkgreen]{hyperref} 

\newtheorem{conj}{Conjecture}
\newcommand{\ie}{i.e. }
\newcommand{\mydot}{\boldsymbol{\cdot}}
\newcommand{\grad}{\bnabla}

\newcommand{\mach}{{\mathcal{M}}}
\newcommand{\mj}{{\mathcal{J}}}
\newcommand{\shom}{s^{(\rm{ho})}}
\newcommand{\sinho}{s^{(\rm{in})}}
\newcommand{\stot}{s}

    \shorttitle{Janzen-Rayleigh inferences}
    \shortauthor{I. S. Wallerstein and U. Keshet}
    \title{
    Compressible potential flows around round bodies:
    Janzen-Rayleigh expansion inferences\\
    }
    \author{Idan S. Wallerstein
    \corresp{\email{wallersh@post.bgu.ac.il}}
    \and Uri Keshet\corresp{\email{ukeshet@bgu.ac.il}}}
    \affiliation{Physics Department, Ben-Gurion University of the Negev, POB 653, Be'er-Sheva 84105, Israel}

\begin{document}

\maketitle

\begin{abstract}
The subsonic, compressible, potential flow around a hypersphere can be derived using the Janzen-Rayleigh expansion (JRE) of the flow potential in even powers of the incident Mach number $\mach_\infty$.
JREs were carried out with terms polynomial in the inverse radius $r^{-1}$ to high orders in two dimensions (2D), but were limited to order $\mach_\infty^4$ in three dimensions (3D).
We derive general JRE formulae to arbitrary order, adiabatic index, and dimension.
We find that powers of $\ln(r)$ can creep into the expansion, and are essential in 3D beyond order $\mach_\infty^4$.
Such terms are apparently absent in the 2D disk, as we confirm up to order $\mach_\infty^{100}$, although they do show in other dimensions (e.g. at order $\mach_\infty^2$ in 4D) and in non-circular 2D bodies.
This suggests that the disk, which was extensively used to study basic flow properties, has additional symmetry.
Our results are used to improve the hodograph-based approximation for the flow in front of a sphere.
The symmetry-axis velocity profiles of axisymmetric flows around different prolate spheroids are approximately related to each other by a simple, Mach-independent scaling.
\end{abstract}

\section{Introduction}
\label{sec:Introduction}

While ideal flows \citep[inviscid, with no heat conduction or additional energy dissipation effects;][]{LandauLifshitz59_FluidMechanics} are an extreme limit, they play an important role in research, for example (i) as a basis for more realistic flows, with additional effects such as viscosity;
(ii) for modeling the bulk of weakly-interacting Bose-Einstein condensate (BEC) superfluids, which can be approximated as an inviscid, compressible fluid with a polytropic index $\gamma=2$;
(iii) for modeling flow regimes which are not sensitive to the level of weak viscosity, such as in front of a round object;
and (iv) for code validation and pedagogical reasons.

Janzen-Rayleigh expansions (JREs) can be broadly identified as an expansion of the flow variables in terms of the Mach number.
JREs were used by \citet{Janzen_1913} and \citet{rayleigh_1916} as a method to study D'Alembert paradox with the addition of compressibility effects.
They considered an inviscid, compressible, subsonic flow of a fluid with a polytropic equation of state (EoS), with no external forces or initial vorticity.
Specifically, a steady flow was assumed around a disk in two dimensions (2D) or around a sphere in three dimensions (3D), with an incident uniform flow far from the body.
Introducing the flow potential, a scalar non-linear partial differential equation (PDE) was obtained, and expanded in the incident Mach number squared, $\mach_\infty^2$.

Using the same setup, the JRE was used to solve the flow around various blunt objects  \citep{Goldstein_Lighthill_1944, Hasimoto_1951, Hida_1953, longhorn_1954, Imai1957, kaplan1957, Van_Dyke_1958}.
The JRE has been generalized to several areas of research, such as vortex flows \citep*[\eg][]{Moore_Pullin_1991,moore_pullin_1998,meiron_moore_pullin_2000,Leppington06,crowdy_krishnamurthy_2018}, porous channel flows \citep*{Majdalani2006,maicke_majdalani_2008, Maicke_Saad_Majdalani2010,Cecil_Majdalani_2015}, and acoustics \citep*{SLIMON2000377,MOON2013}, and could be used in a wide range of other applications, as we demonstrate below.

The 2D flow around a disk has been researched extensively \citep{Janzen_1913,rayleigh_1916,VanDykeGuttmann83,Guttmann_and_Thompson}, for example in search for a solution to the transonic controversy, namely, ``the existence, or non-existence, of a continuous transonic flow, that is, without a shock wave, around a symmetrical wing profile, with zero incidence with respect to the undisturbed velocity'' \citep{ferrari1966}.
These and other problems that require a high-order expansion could not be explored as thoroughly in the 3D case, because previous JREs for the sphere were limited to second, \ie $\mach^4_\infty$, order \citep{kaplan1940flow, Tamada}.
Indeed, a power series in $r^{-1}$ yields a non-physical behaviour at the third, \ie $\mach^6_\infty$, expansion order.
A higher order expansion in 3D is also needed, for example, to derive the flow in front of a sphere, in order to model axisymmetric bodies in various fields of physics \citep[\eg][and references therein]{Keshet_Naor}.
Flows in higher dimensions, $d>3$, are also important, mainly for theoretical and pedagogical purposes.

We study the steady, inviscid, compressible flow around a hypersphere in $d$ spatial dimensions, and provide explicit results for the $d={2}$ (disk) and $d=3$ (sphere) cases.
We derive the JRE in 3D beyond the presently available second-order, to an arbitrary order.
As an example of the usage of such an expansion, we compare the axisymmetric flow in front of a sphere to that in front of a spheroid, and show that a simple scaling approximates the flow well for the prolate case.
Furthermore, three orders in the JRE are sufficient to adequately describe these flows, for Mach numbers ranging from the incompressible to the sonic regimes.

The paper is organized as follows.
In \S\ref{sec:Janzen-Rayleigh Expansion}, we derive the JRE equations from the hydrodynamical ones.
In \S\ref{sec:JRE for a hypersphere}, we show that each term in the JRE of the flow potential around a hypersphere is a finite sum of a product of powers of the radial coordinate, powers of its logarithm, and a set of orthogonal functions (Jacobi polynomials) of the polar coordinate.
In \S\ref{sec:Semi-analytical and numerical solvers for disk and sphere flows}, we outline the semi-analytic algorithm to compute the JRE and a numerical pseudospectral method we use to solve the non-linear compressible flow.
We present results from our numerical solver and from the semi-analytical JRE, and highlight the compressibility to compute the JRE, in \S\ref{sec:results}.
The JRE is used to improve a hodographic approximation for the flow in front of the sphere in \S\ref{sec:Application example: axial hodographic approximation}.
In \S\ref{sec:Flow around spheroids}, we show that the axisymmetric flow in front of prolate spheroids is well approximated by that of the scaled flow in front of a sphere.
We summarize and discuss our results in \S\ref{sec:Summary and discussion}.
In appendix \S\ref{sec:Explicit JRE coefficients for low orders}, we provide explicit values of the coefficients of the series representation of the JRE
for low orders of the flows around a disk and a sphere. Appendix \S\ref{sec:Hodographic approximation of the solution of radialflow}
discusses the hodographic approximation and general results. In appendix \S\ref{sec:details and convergence of the pseudospectral solver} we detail the numerical solver.

\section{Janzen-Rayleigh Expansion}
\label{sec:Janzen-Rayleigh Expansion}

Consider an isentropic flow in $d$-dimensions with no external forces of a perfect fluid with a polytopic, ideal gas EoS of adiabatic index $\gamma$,
\begin{equation}
    \label{eq: equation of state}
    p \propto {\rho ^\gamma } \,.
\end{equation}
The equations governing the flow are the continuity equation,
\begin{equation}
    \label{eq: mass continuity equation} \frac{{\partial \rho }}{{\partial t}} + \grad \mydot \left( {\rho \bm{u}} \right) = 0 \, ,
\end{equation}
and the Euler equation,
\begin{equation}
    \label{eq: Euler momentum equation}
    \frac{{\partial {\bm{u}}}}{{\partial t}} + ({\bm{u}} \mydot \grad) {\bm{u}} =  - \frac{{\grad p}}{\rho } = -\frac{\grad c^2}{\gamma-1}\, .
\end{equation}
Here, $\rho$, ${\bm u}$, and $p$ are respectively the mass density, velocity, and pressure,
$c\equiv(dp/d\rho)^{1/2}=(\gamma p/\rho)^{1/2}$ is the sound velocity, and $\grad$ is the del operator in $d$-dimensions.

We henceforth assume a steady flow.
Combining (\ref{eq: equation of state}) and (\ref{eq: mass continuity equation}) to eliminate $\rho$ in favor of $c$ then yields
\begin{equation}
    \label{eq: modified continuity}
    \frac{1}{\gamma-1}{\bm u} \mydot \grad {c^2} =  -{c^2}\grad  \mydot {\bm u} \, .
\end{equation}
For our inviscid, steady flow, (\ref{eq: Euler momentum equation}) yields the Bernoulli principle, namely, the quantity $w{\bm{u}}^2+c^2$ is constant along streamlines, where we define $w\equiv(\gamma-1)/2$.

We consider the flow along a body for a uniform flow incident from infinity,
\begin{equation}
    \label{eq: boundary conditions at infinity}
    c\left( {r \to \infty } \right) = {c_{\infty}} \quad \mbox{and}\quad
    {\bm{u}}\left( {r \to \infty } \right) = u_{\infty}\bm{\hat z} \, ,
\end{equation}
where $r$ is the radial coordinate, $z$ is the coordinates along the flow, and the subscript $\infty$ indicates the far region, tending to spatial infinity.
Using the boundary conditions (\ref{eq: boundary conditions at infinity}) and the assumption that every streamline starts at infinity, the Bernoulli equation may be written as
\begin{equation}
    \label{eq: Bernoulli principle}
    w{{\bm u}^2} + c^2  = w{{\bm u}_\infty ^2} + c_\infty ^2 \,.
\end{equation}

Considering the subsonic regime, we henceforth restrict the discussion to a potential flow, writing the velocity as the gradient of the flow potential,
\begin{equation}
    \label{eq: flow potential definition}
    \bm{u} \equiv \grad \phi \, .
\end{equation}
Isolating $c^2$ from (\ref{eq: Bernoulli principle}), substituting it into (\ref{eq: modified continuity}), and using the potential (\ref{eq: flow potential definition}), we obtain a single PDE for the potential $\phi$ \citep{rayleigh_1916},
\begin{equation}
\label{eq: flow potential non-linear PDE}
    \frac{1}{2}(\grad \phi)  \mydot \grad {\left( {\grad \phi } \right)^2} = \left[ w {u_\infty ^2} + c_\infty ^2 - w {{{\left( {\grad \phi } \right)}^2}} \right]{\nabla ^2}\phi \,.
\end{equation}
We normalize the variables to obtain them in a dimensionless form, by taking $\phi\to (u_{\infty}R)\phi$, and ${\bm r} \to R{\bm r}$ (which also normalize the velocity ${\bm{u}}\to u_\infty{\bm{u}}$), where $R$ is a characteristic length scale, for example the radius of a hypersphere.
Defining $\mach_{\infty} \equiv u_{\infty}/c_{\infty}$
as the Mach number at infinity, we then arrive at the dimensionless
\begin{equation}
    \label{eq: normalized flow potential non-linear PDE}
    \frac{1}{2}(\grad \phi)  \mydot \grad {\left( {\grad \phi } \right)^2} = \left[ {w + {\mach_\infty ^{-2}} - w{\left( {\grad \phi } \right)}^2} \right]{\nabla ^2}\phi \,.
\end{equation}

We supplement this equation for the potential with two boundary conditions (BCs): the uniform incident flow from infinity, and the slip, no-penetration condition on the surface of the body (${\bm{\hat n}}\mydot {\bm{u}}=0$), namely
\begin{equation}
    \label{eq: flow potential boundary conditions}
    \phi \left( {r \to \infty } \right) = z = r\cos \theta \quad\mbox{and}\quad
    \bm{\hat n}\mydot \grad \phi  = 0 \,.
\end{equation}
Here, $\bm{\hat n}$ is the normal to the body, and we use hyperspherical coordinates: a radial coordinate $0 \le r \le \infty$, a polar angle $0 \le \theta \le \pi$ measured with respect to the uniform flow at infinity, and $(d-2)$ additional angles.
For axisymmetric flows such as in the case of a hypersphere, the flow is by symmetry independent of these additional angles.

To arrive at the JRE, we substitute the expansion
\begin{equation}
    \label{eq: JRE for flow potential}
    \phi{(\bm r)}  \equiv \sum\limits_{m = 0}^\infty {{\phi _m(\bm r)}{\mach_{\infty}^{2m}}}
\end{equation}
into (\ref{eq: normalized flow potential non-linear PDE}), and isolate the different powers of $\mach_\infty$.
This leads to a recursive PDE for $\phi_m$ (for $m \geq 1$),
\begin{eqnarray}
      \label{eq: JRE flow potential linear PDE}
      {\nabla ^2}{\phi _m} & = & - w{\nabla ^2}{\phi _{m - 1}} \\
      & & +w \sum\limits_{\{{m_1},{m_2},{m_3}\} = 0}^{m}  {\delta_{{m_1} + {m_2} + {m_3},m - 1} \left( {\grad {\phi _{{m_1}}} \mydot \grad {\phi _{{m_2}}}} \right){\nabla ^2}{\phi _{{m_3}}}} \nonumber
      \\
      & & +\frac{1}{2} \sum\limits_{\{{m_1},{m_2},{m_3}\} = 0}^{m} {\delta_{{m_1} + {m_2} + {m_3},m - 1} \grad \left( {\grad {\phi _{{m_1}}} \mydot \grad {\phi _{{m_2}}}} \right) \mydot \grad {\phi _{{m_3}}}} \nonumber\,,
\end{eqnarray}
with an initial equation
\begin{equation}
      \label{eq: JRE PDE m=0}
      {\nabla ^2}{\phi _0}  = 0\,,
\end{equation}
along with BCs
\begin{equation}
\label{eq: BC for phi_m}
        {\phi _m}\left( {r \to \infty } \right) = {\delta _{m,0}} r\cos \theta
\end{equation}
at spatial infinity, and
\begin{equation}
\label{eq: BC for phi_m_body}
        {\bm{\hat n}} \mydot \grad {\phi _m}  = 0
\end{equation}
on the body.
Here, $\delta_{i,j}$ is the Kronecker delta symbol.

The zeroth-order equation (\ref{eq: JRE PDE m=0}) is the Laplace equation for $\phi_0$, corresponding to the incompressible limit $\mach_\infty\to 0$.
For higher orders, equation (\ref{eq: JRE flow potential linear PDE}) can be regarded as a set of Poisson equations for $\phi_m$ at each order $m$, with a source term being a function of the lower-order solutions, $\{\phi_i\}^{m-1}_{i=0}$, and their derivatives.
In general, the solution to (\ref{eq: JRE flow potential linear PDE}) at any order $m\geq1$ under the BCs (\ref{eq: BC for phi_m}) and (\ref{eq: BC for phi_m_body}) is a sum of an inhomogeneous solution and a homogeneous solution,
\begin{equation}
\label{eq: potential sum of in and homogeneous solutions}
    \phi_m = \phi_m^{\rm{(in)}} + \phi_m^{\rm{(ho)}},
\end{equation}
where $\phi_m^{\rm{(ho)}}$ solves the Laplace equation.

The JRE (\ref{eq: JRE for flow potential}) is constructed by iteratively determining the functions $\phi_m$.
One starts by deriving $\phi_0$, obtained as the solution to the zeroth-order Laplace equation (\ref{eq: JRE PDE m=0}) under the BCs.
Increasingly higher-order functions $\phi_m$ are then incrementally derived, by solving the corresponding Poisson equations (\ref{eq: JRE flow potential linear PDE}) under the same BCs.
At each order $m\geq1$, the inhomogeneous solution $\phi_m^{\rm{(in)}}$ is fixed by the source term in the corresponding (\ref{eq: JRE flow potential linear PDE}), which is explicitly written once all lower order functions $\phi_m$ are determined.
This solution is then combined with an expansion $\phi_m^{\rm{(ho)}}$ of homogenous solutions, the coefficients of which are determined by applying the BCs to the resulting (\ref{eq: potential sum of in and homogeneous solutions}).

\section{JRE for a hypersphere}
\label{sec:JRE for a hypersphere}

Simple solutions exist for the flow around a $d$-dimensional hypersphere, for which the Laplace equation has known analytic solutions, and the Poisson equation is easily solved.
Choosing the characteristic length $R$ as the hypersphere radius, the normalized slip BC (\ref{eq: BC for phi_m_body}) here simplifies to
\begin{equation}
    \label{eq: hypersphere slip bc}
    \partial_r\phi\left(r=1,\theta\right) = 0\,.
\end{equation}

Considering the hyperspherical and axial symmetries, it is useful to write the general solution to the Laplace equation (\ref{eq: JRE PDE m=0}) as the infinite sum of positive and negative radial powers \citep{Feng_2011}
\begin{equation}
    \label{eq: laplace general solution}
    \sum\limits_{n = 1}^\infty {\left(A_n r^{n}+B_n r^{ - n - d + 2}\right)\mj_n^{(d)}\left(\mu\right)},
\end{equation}
where we define $\mu \equiv \cos \theta$, the normalized Jacobi polynomials
\begin{equation}
    \label{eq: normalized Jacobi polynomials definition}
    \mj_n^{(d)}\left( {\mu } \right) \equiv \frac{{J_n^{\left( {\frac{{d - 3}}{2},\frac{{d - 3}}{2}} \right)}\left( {\mu } \right)}}{{J_n^{\left( {\frac{{d - 3}}{2},\frac{{d - 3}}{2}} \right)}\left( 1 \right)}}
    = \frac{C_n^{\left(\frac{d}{2}-1\right)}(\mu)}{{{n+d-3}\choose {n}}}
    \,,
\end{equation}
and numerical coefficients $A_n$ and $B_n$.
Here, $J_n^{(\alpha,\beta)}(\mu)$ and $C_n^{(\alpha)}(\mu)$ are respectively the standard Jacobi and Gegenbauer polynomials.

The source term in (\ref{eq: JRE flow potential linear PDE}) is a sum of multiple terms, each composed of even derivatives in $\theta$ and odd multiples of functions $\phi$.
Therefore, the polar dependence will always exhibit the same symmetry as the zero-order solution $\phi_0$.
As the hypersphere is isotropic, the incompressible flow also shows a backward-forward symmetry.
We conclude that the functions $\mj_n^{(d)}$ appearing in $\phi_m$ on all orders $m$ have only odd $n$.
Next, we construct the hypersphere JRE by considering increasingly larger orders $m$.

\subsection{Zeroth order: $m=0$}
\label{subsec: Zeroth order: $m=0$}
For the $m=0$ order, the infinity BC (\ref{eq: BC for phi_m}) allows only the radially linear and decaying terms in the solution (\ref{eq: laplace general solution}).
The slip BC (\ref{eq: hypersphere slip bc}) restricts the solution further, allowing for only the decaying term proportional to $\mj_1^{(d)}(\mu) = \mu$.
The solution to the Laplace equation (\ref{eq: JRE PDE m=0}) around a $d$-dimensional hypersphere thus becomes
\begin{equation}
    \label{eq: flow potential m=0 solution}
    \phi _0^{(d)}(r,\theta) =
    \left[ {r + \frac{1}{{\left( {d - 1} \right){r^{d - 1}}}}} \right]\cos\theta\,.
\end{equation}
Henceforth, we omit the dimension superscript $(d)$ unless necessary.

\subsection{First order: $m=1$}
The first-order potential, $\phi_1$, is determined by
\begin{eqnarray}
\label{eq: JRE PDE m=1}
    {\nabla ^2}{\phi _1} & = & - w{\nabla ^2}{\phi _{0}} +\frac{1}{2} \left( \grad {\phi _{{0}}} \right) \mydot \grad \left( \grad {\phi _{{0}}} \right)^2  +w \left( \grad {\phi _{{0}}} \right)^2{\nabla ^2}{\phi _{0}}\,,
\end{eqnarray}
which, by the zeroth-order (\ref{eq: JRE PDE m=0}), simplifies  to
\begin{eqnarray}
    \label{eq: JRE PDE m=1 simplified}
    {\nabla ^2}{\phi _1} & = & \frac{1}{2} \left( \grad {\phi _{{0}}} \right) \mydot \grad \left( \grad {\phi _{{0}}} \right)^2.
\end{eqnarray}
Plugging the $m=0$ solution (\ref{eq: flow potential m=0 solution}) into (\ref{eq: JRE PDE m=1 simplified}) gives the Laplace equation with the explicit source term,
\begin{eqnarray}
\label{eq: JRE PDE m=1 explicit not linearized}
    {\nabla ^2}{\phi _1} & = & \left[-3\mu+\left(2+d\right)\mu^3\right]r^{-d-1}\frac{d }{d-1} + \left[\left(-3+d\right)\mu+\left(2+d-d^2\right)\mu^3\right]r^{-2d-1}\frac{2d}{\left(d-1\right)^2} \nonumber \\
    & & +\left[\left(-3+2d\right)\mu+\left(2+d-3d^2+d^3\right)\mu^3\right]r^{-3d-1}\frac{d}{\left(d-1\right)^3} \,.
\end{eqnarray}
The orthogonal decomposition of Jacobi polynomials \citep{CHAGGARA2010609} then yields
\begin{eqnarray}
\label{eq: JRE PDE m=1 explicit}
    {\nabla ^2}{\phi _1} & = &  \mj_3\left( {\mu } \right)r^{-d-1} d\\
    &+& \left[-2d \mj_1\left( {\mu } \right) -\left(1+d\right)\left(d-2\right)\mj_3\left( {\mu } \right)\right] r^{-2d-1} \frac{2d }{\left(d-1\right)\left(d+2\right)} \nonumber \\
    &+& \left[d\left(-4+3d\right) \mj_1\left( {\mu } \right) -\left(1+d-d^2\right)\left(d-2\right)\mj_3\left( {\mu } \right)\right] r^{-3d-1} \frac{d }{\left(d-1\right)^2\left(d+2\right)} \nonumber \, .
\end{eqnarray}
This result may be compactly written in the form
\begin{equation}
\label{eq: JRE PDE m=1 explicit sum}
    {\nabla ^2}\phi_1 =  \sum\limits_{k,n}^{} {\sinho_{1,k,n}{r^{k}}\mj_n\left( {\mu } \right)} \,,
\end{equation}
where $\sinho_{m,k,n}$ are the expansion coefficients of the order-$m$ Poisson source term, for radial order $k$ and angular order $n$.

As the Poisson equation is linear, suffice to solve, for arbitrary $k$ and $n$, the equation
\begin{equation}
    \label{eq: JRE PDE genral Poisson source no logs}
    {\nabla ^2}\phi_{k,n} = {r^{k}}\mj_n\left( {\mu } \right) \,.
\end{equation}
This equation has the particular, inhomogeneous solution
\begin{equation}
\label{eq: JRE PDE inhomogeneous solution no logs}
    \phi _{k,n}^{(\rm{in})} = \frac{{{r^{k+2}}\mj_n\left( {\mu } \right)}}{{(k+2) \left( {k + d} \right) - n\left( {n + d - 2} \right)}} \, ,
\end{equation}
provided that $k \notin \{- d - n, n-2\}$.
These two exceptional values of $k$ do not occur at order $m=1$, as seen from (\ref{eq: JRE PDE m=1 explicit}), but they may appear at higher orders, as discussed below in \S\ref{Higher orders: m>=2}.
We may now expand the inhomogeneous solution as
\begin{equation}
\label{eq: JRE PDE inhomogeneous solution sum}
 \phi^{\rm{(in)}}_1 = \sum\limits_{k,n}{\sinho_{1,k,n}\phi^{\rm{(in)}}_{k,n}} \,,
\end{equation}
where the numerical coefficients $\sinho_{1,k,n}$ are determined by equating (\ref{eq: JRE PDE m=1 explicit}) and (\ref{eq: JRE PDE m=1 explicit sum}).

From (\ref{eq: JRE PDE m=1 explicit}) and (\ref{eq: JRE PDE inhomogeneous solution no logs}) we see that the largest (\ie least negative) power of $r$ for $m=1$ is $k_{max}=-d+1<0$, implying that $\phi^{\rm{(in)}}_1$ includes only radially declining terms.
For higher orders, (\ref{eq: JRE flow potential linear PDE}) combines the derivatives of multiple $\phi_m$ functions, but the highest radial power $k_{max}$ in the source term remains no larger than $-d-1$.
Consequently, the inhomogeneous solution for all $m\geq1$ orders includes only radially declining terms.
This conclusion, combined with the BCs, indicate that the homogeneous solution may be expanded with only negative powers of $r$ for any $m>0$,
\begin{equation}
    \label{eq: JRE PDE homogeneous solution}
    \phi^{\rm{(ho)}}_{m>0} = \sum\limits_{n = 1}^\infty  {\shom_{m,n} {r^{ - n - d + 2}}\mj_n\left(\mu\right)} \,,
\end{equation}
where $\shom_{m,n}$ are numerical coefficients, to be determined below.

The full solution for $m=1$ now becomes
\begin{equation}
\label{eq: JRE PDE m=1 total solution}
    \phi_1 = \phi_1^{\rm{(in)}} + \phi_1^{\rm{(ho)}}  = \sum\limits_{k,n}{\stot_{1,k,n} r^k \mj_n\left(\mu\right)},
\end{equation}
where we introduced the numerical coefficients
\begin{equation}
\label{eq: JRE PDE m=1 total solution coefficient}
    \stot_{1,k,n}=\frac{\sinho_{1,k-2,n}}{k(k+d-2)-n(n+d-2)}+\shom_{1,n}\delta_{k,-n-d+2} \,.
\end{equation}
These coefficients may now be determined from the slip BC (\ref{eq: hypersphere slip bc}),
\begin{equation}
\label{eq: JRE potential m=1 body BC}
    \frac{\partial \phi_1}{\partial r}(r=1) = \sum\limits_{k,n}{k \stot_{1,k,n} \mj_n\left(\mu\right)} = 0\,,
\end{equation}
implying that
\begin{equation}
\label{eq: JRE PDE homoegeneous coefficient}
    \shom_{1,n} = \frac{1}{n+d-2} \sum\limits_{k,n}{\frac{k \sinho_{1,k-2,n}}{k(k+d-2)-n(n+d-2)}}\,.
\end{equation}
As the coefficients $\sinho_{1,k,n}$ are known from (\ref{eq: JRE PDE m=1 explicit}) and (\ref{eq: JRE PDE m=1 explicit sum}), the solution (\ref{eq: JRE PDE m=1 total solution}) is completely specified by (\ref{eq: JRE PDE m=1 total solution coefficient}) and (\ref{eq: JRE PDE homoegeneous coefficient}).

\subsection{Higher orders: $m\geq2$}
\label{Higher orders: m>=2}
Substituting $\phi_0$ and $\phi_1$ in the Poisson equation (\ref{eq: JRE flow potential linear PDE}) for the next, $m=2$ order, results again in an equation of the form (\ref{eq: JRE PDE m=1 explicit sum}), but now for $\phi_2$ and with different coefficients.
Equations of the same form persist for increasingly higher orders $m$, as long as the source term in (\ref{eq: JRE flow potential linear PDE}) is free of non-zero terms $\stot_{m,k,n}$ which satisfy one of the aforementioned special conditions, $k=- d - n$ or $k=n-2$.
Indeed, the source term consists of differential operators of the form $\nabla^2f(r,\mu)$ and $\grad f(r,\mu)\cdot\grad g(r,\mu)$, where $f$ and $g$ are constructed from lower order $\phi$ terms.
As long as $f$ and $g$ are sums of terms, each of which is a product of powers of $r$ and Jacobi polynomials in $\mu$, the source term would remain of the form (\ref{eq: JRE PDE m=1 explicit sum}).

For the special cases $k \in \{- d - n, n-2\}$, the solution to (\ref{eq: JRE PDE genral Poisson source no logs}) is no longer given by (\ref{eq: JRE PDE inhomogeneous solution no logs}), which formally diverges.
It is sufficient to consider the former case, $k=-d-n$, as the latter, $k = n -2$, is then obtained by the transformation $n \to - n - d + 2$, under which $\mj_n^{(d)}$ remains invariant.
Here, the inhomogeneous solution to (\ref{eq: JRE PDE genral Poisson source no logs}) becomes
\begin{equation}
\label{eq: potential first solution with logs}
    \phi_{k=-d-n,n} =  -\frac{{\Gamma \left[ {2,\left( {2n + d - 2} \right)\ln \left( r \right)} \right]}}{{{{\left( {2n + d - 2} \right)}^2}}} {r^n} \mj_n\left( {\mu } \right) \,,
\end{equation}
where $\Gamma(x,y)$ is the incomplete Gamma function, defined by
\begin{equation}
\label{eq: Incomplete Gamma function definition}
    \Gamma \left( {x,y} \right) = \int\limits_y^\infty  {{t^{x - 1}}{e^{ - t}}dt} \,.
\end{equation}

Once a $\ln(r)$ term of (\ref{eq: potential first solution with logs}) appears in the inhomogeneous solution for $\phi_m$, as a result of a special $k$ term in the source-term expansion (\ref{eq: JRE PDE m=1 explicit sum}), a power-series in $r$ as in (\ref{eq: JRE PDE m=1 total solution}) is no longer sufficient.
Indeed, when $x=\ell+1$ and $y=p\ln \left(r\right)$ for integers $\ell$ and $p$,
\begin{equation}
\label{eq: Incomplete Gamma function property}
    \Gamma \left[ {\ell + 1,p\ln \left(r\right)} \right] = \frac{{\ell!}}{{{r^p}}}\sum\limits_{j = 0}^\ell {\frac{{{{\left[p\ln \left(r\right) \right]}^j}}}{{j!}}}
\end{equation}
contains logarithmic, and not only polynomial, radial terms.
Since $\ln(r)$ cannot be expanded as a power series of $r^{-1}$ valid over the full $1\leq r<\infty$ range, one must then take into account logarithmic terms in the expansion of $\phi_m$ and in the resulting source terms of higher order functions.

Generalizing the prototypical source term in (\ref{eq: JRE PDE m=1 explicit sum}) to include a logarithm of $r$ to some power $l$, we must therefore also consider the generalized equation
 \begin{equation}
\label{eq: JRE PDE genral Poisson source with logs}
    {\nabla ^2}\phi_{k,n,\ell} = r^{k} \mj_n\left( {\mu } \right) \ln^\ell\left(r\right) \,.
 \end{equation}
As long as $k$, $d$, and $n$ do not satisfy one of the special cases
\begin{equation}\label{eq:BadK}
k \in \{ - d - n, n-2\} \, ,
\end{equation}
the inhomogeneous solution to this equation is
\begin{eqnarray}
\label{eq: JRE PDE inhomogeneous solution with logs}
    \phi_{k,n,\ell} & = & \frac{{\mj_n\left( {\mu } \right)}}{{2n + d - 2}} \left({r^{-n-d+2}}\frac{{\Gamma \left[ {\ell + 1, \left( {-d - k - n } \right)\ln \left(r\right)} \right]}}{{{{\left( { - d - k - n} \right)}^{\ell + 1}}}} \right. \nonumber\\
    &&  \left. - {r^n}\frac{{\Gamma \left[ {\ell + 1,\left( { n - k -2} \right)\ln \left(r\right)} \right]}}{{{{\left({ n - k-2} \right)}^{\ell + 1}}}} \right) \,,
\end{eqnarray}
whereas for $k=-d-n$ we find
\begin{equation}
\label{eq: JRE PDE inhomogeneous solution with logs special}
        \phi _{k=- d - n,n,\ell} = - {r^n}\mj_n\left( {\mu } \right)
        \frac{{\Gamma\left[ {\ell + 2,\left( {2n + d - 2} \right)\ln \left(r\right)} \right]}}{{{{\left(\ell + 1\right)\left( {2n + d - 2} \right)}^{\ell + 2}}}} \,.
\end{equation}
The case $k=n-2$ is again obtained using the transformation $n \to - n - d + 2$.

\begin{equation}
\label{eq: JRE PDE inhomogeneous solution with logs special B}
        \phi _{k=-n-d+2,n,\ell} = - {r^{-n-d+2}}\mj_{n}
        \left( {\mu } \right)
        \frac{{\Gamma\left[ {\ell + 2,\left( {-2n-d+2} \right)\ln \left(r\right)} \right]}}{{{{\left(\ell + 1\right)\left( {-2n-d+2} \right)}^{\ell + 2}}}} \,.
\end{equation}

The overall solution for $\phi_m$ may now be expanded as the finite sum
\begin{equation}
\label{eq: JRE PDE m>2 total solution}
    \phi_m = \sum\limits_{k,n,\ell}{\stot_{m,k,n,\ell}r^k\mj_n^{(d)}\left( {\mu } \right)\ln^{\ell} \left(r\right)}\,,
\end{equation}
where the numerical coefficients $\stot_{m,k,n,\ell}$ are determined using the BCs in analogy with the above $m=2$ discussion.
The expansion (\ref{eq: JRE PDE m>2 total solution}) is complete, as the inhomogeneous solution yields only powers of $r$ and $\ln \left(r\right)$, and the homogeneous solution yields only powers of $r$, so the resulting higher-order source terms are again of the form (\ref{eq: JRE PDE genral Poisson source with logs}).

One can prove, by induction, the following rules for the $m\geq1$ indices in (\ref{eq: JRE PDE m>2 total solution}),
\begin{equation}
1-d-2md \leq k \leq 1-d
\end{equation}
and
\begin{equation}
1\leq n \leq 1+2m\, ;
\end{equation}
the summation limits for the logarithmic term are discussed in \S\ref{sec:results}.

\section{Semi-analytical and numerical solvers for disk and sphere flows}
\label{sec:Semi-analytical and numerical solvers for disk and sphere flows}

We explicitly solve equation (\ref{eq: JRE flow potential linear PDE}) for the hypersphere in the $d=2$ and $d=3$ cases, namely, we derive the flows around a 2D disk and around a 3D sphere.
The normalized Jacobi functions reduce to the Chebyshev polynomials ${T_n}$ in 2D,
\begin{equation}
\label{eq: normalized Jacobi polynomials in 2D and Chebyshev polynomials definition}
        \mj_n^{(2)}\left( {\cos \theta } \right) = {T_n}\left( {\cos \theta } \right) = \cos \left( {n\theta } \right) \, ,
\end{equation}
and to the Legendre polynomials ${P_n}$ in 3D,
\begin{equation}
\label{eq: normalized Jacobi polynomials in 2D and Legendre polynomials}
        \mj_n^{(3)}\left( {\cos \theta } \right) = {P_n}\left( {\cos \theta } \right) \, .
\end{equation}
As discussed in \S\ref{sec:JRE for a hypersphere}, we consider the generalized expansion (\ref{eq: JRE for flow potential}) and (\ref{eq: JRE PDE m>2 total solution}) of the flow potential in each case,
\begin{equation}
\label{eq: JRE for potential 2D around disk}
    \phi^{\rm{disk}} = \left(r+\frac{1}{r}\right) \cos\theta + \sum\limits_{m=1}^{\infty}
    {\sum\limits_{\ell=0}^{\infty}
    {\sum\limits_{n=1}^{2m+1}
    {\sum\limits_{k=-1-4m}^{-1}
    {a_{m,k,n,v}{\mach_\infty^{2m}}{r^k}\cos \left( {n\theta } \right){{\ln }^\ell}\left(r\right)}}}}
\end{equation}
and
\begin{equation}
\label{eq: JRE for potential 3D around sphere}
    \phi^{\rm{sphere}} = \left(r+\frac{1}{2r^2}\right) \mu + \sum\limits_{m=1}^{\infty}
    {\sum\limits_{\ell=0}^{\infty}
    {\sum\limits_{n=1}^{2m+1}
    {\sum\limits_{k=-2-6m}^{-2}
    {{b_{m,k,n,\ell}}{\mach_\infty^{2m}}{r^k}{P_{n}}\left(\mu\right){{\ln }^\ell}\left(r\right)}}}} \,,
\end{equation}
where $a$ and $b$ are the expansion coefficients in 2D and 3D, respectively and the summation index $n$ takes only odd values.

We compute these JREs analytically, following the steps outlined in \S\ref{sec:JRE for a hypersphere}.
The $m=0$ term is obtained from (\ref{eq: flow potential m=0 solution}).
For each order $m>0$, we compute the source terms on the RHS of (\ref{eq: JRE flow potential linear PDE}), and decompose them into a sum of terms proportional to $r^{k} \mj_n( {\mu } ) \ln^\ell(r)$ as in (\ref{eq: JRE PDE genral Poisson source with logs}).
The inhomogeneous solution $\phi^{\rm{(in)}}_m$ is then obtained as the corresponding sum of solutions of the form (\ref{eq: JRE PDE inhomogeneous solution with logs})--(\ref{eq: JRE PDE inhomogeneous solution with logs special B}), and added to the homogeneous solution $\phi^{\rm{(ho)}}_m$ of (\ref{eq: JRE PDE homogeneous solution}).
The coefficients ${\shom_{m,n}}$ in the latter sum are determined from the slip BC, as demonstrated for $m=1$ in (\ref{eq: JRE PDE homoegeneous coefficient}).
This process is repeated until the necessary accuracy of the JRE is reached.

The JRE results are compared to the flow obtained from a numerical solution to the non-linear equation (\ref{eq: normalized flow potential non-linear PDE}).
We follow \citet{PHAM2005203} and use the pseudospectral collocation method \citep{boyd2001chebyshev} for a fast convergence.
In the pseudospectral method, the solution to a PDE is expanded in terms of base functions, each being a product of individual (and typically orthogonal) functions for each coordinate.
In collocation methods, the PDE is then evaluated and solved at a set of points, usually taken as the roots of the basis functions.
This leads to a set of linear equations for the basis function coefficients, which are solved to give an approximate PDE solution.
In general for pseudospectral methods, if the solution is smooth (all derivatives of all orders are continuous) then the convergence is exponential in the number $N$ of collocation points \citep{boyd2001chebyshev}, $O\left(e^{-N}\right)$.

\citet{PHAM2005203} use the Chebyshev-Fourier set of basis functions, invoking $\cos(n\theta)$ and $\sin(n\theta)$ with integer $n$ as angular basis functions.
As discussed in \S\ref{sec:JRE for a hypersphere}, the analytical solution for $\phi$ is a sum of $\cos(n\theta)$ with odd positive $n$, so we expand the flow potential in odd cosine functions.
To achieve the same accuracy as \citet{PHAM2005203}, we need only a fourth of the angular basis functions.

For the radial part, we apply an inversion map $\varrho\equiv r^{-1}$ and expand in Chebyshev Polynomials $T_k(\varrho)$, to resolve both small and large $r$ behavior.
The pseudospectral expansion thus becomes
\begin{equation}
    \label{eq: potential for pseudospectral method}
    {\phi _{\rm{ps}}} = \phi_0 + \sum\limits_{k=0}^{k_{\max}}\sum\limits_{n=0}^{n_{\max}} {{c_{k,n}}{T_k}\left(\varrho\right)\cos \left[ {\left( {2n + 1} \right)\theta } \right]}\, .
\end{equation}
Here, $k_{max}$ and $n_{max}$ are the radial and angular resolution, \ie the number of collocation points, and $c_{k,n}$ are real coefficients.
The BCs are guarantees if they are satisfied by the potential $\phi_0$, chosen as the incompressible solution (\ref{eq: flow potential m=0 solution}).
Since the effective computational domain and symmetry are the same in our 2D and 3D frameworks, we expand both flows, around a disk and around a sphere, in the same basis functions (\ref{eq: potential for pseudospectral method}).
Appendix \S\ref{sec:details and convergence of the pseudospectral solver} provides more details on the the calculations of $c_{k,n}$ and demonstrates the numerical convergence.

\section{Results}
\label{sec:results}

We calculate the JRE (\ref{eq: JRE for potential 2D around disk}) and (\ref{eq: JRE for potential 3D around sphere}) up to order $m=30$ for the disk and $m=18$ for the sphere.
Appendix \S\ref{sec:Explicit JRE coefficients for low orders} provides explicit expressions for the coefficients $a$ and $b$ up to order $m=3$ (\ie $\mach_\infty^6$) for a general $\gamma$.
For the pseudospectral code we use a resolution up to 32 Chebyshev and 64 odd cosine functions, but as shown below, much lower orders and resolutions are sufficient to capture the behaviour of the flow (see figure \ref{fig:spherecompare} and \ref{fig:sphereconv}).
The results are presented in the following figures mainly for $\gamma=7/5$.
Different choices of $\gamma$ give qualitatively similar results, as demonstrated for $\gamma=5/3$ in figure \ref{fig:JRE Hodogrpahic comparison gamma 5/3}.

\begin{figure*}
\centering
    \includegraphics[width=.48\linewidth]{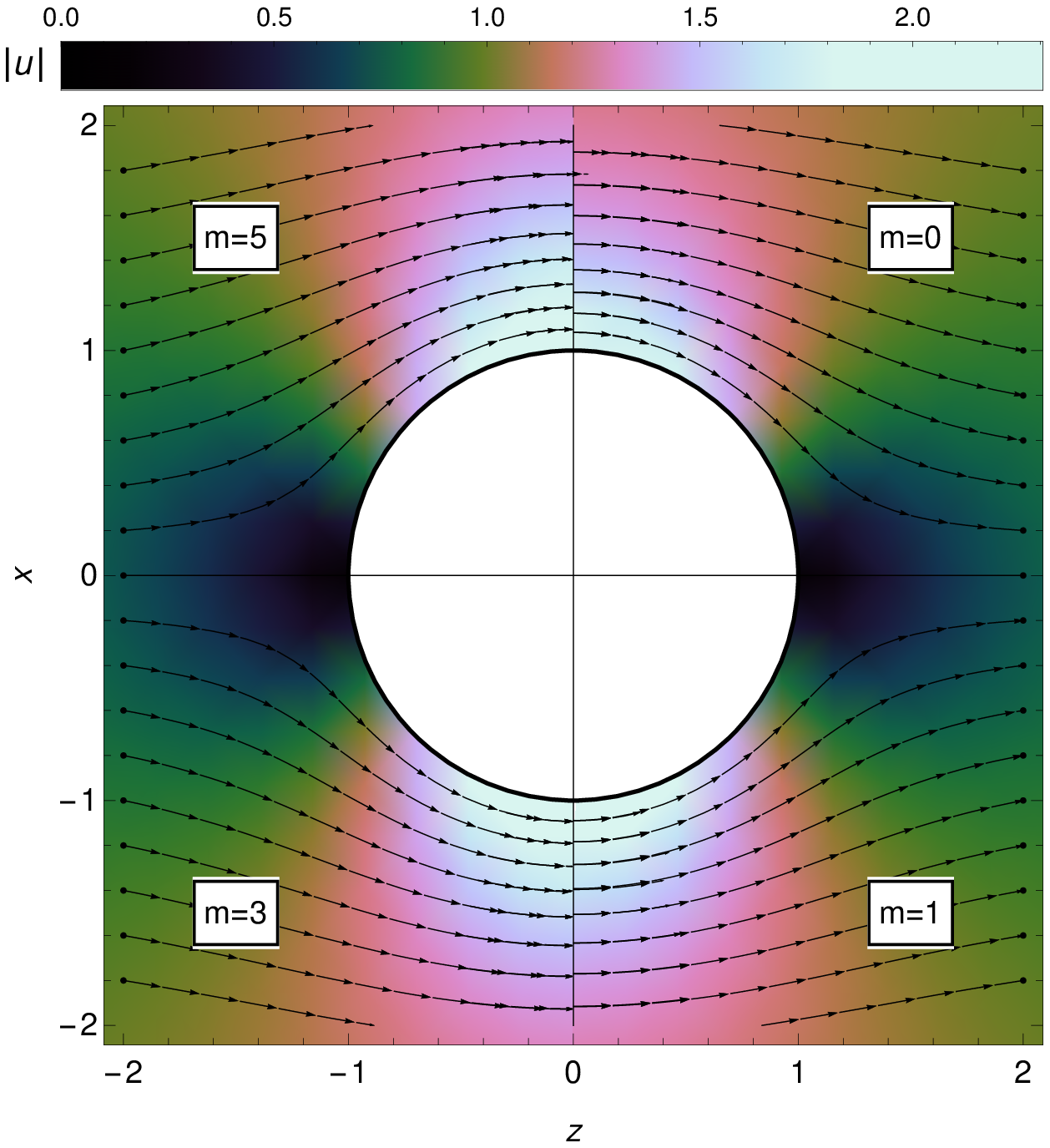}
    \hspace{0.1cm}
    \includegraphics[width=.48\linewidth]{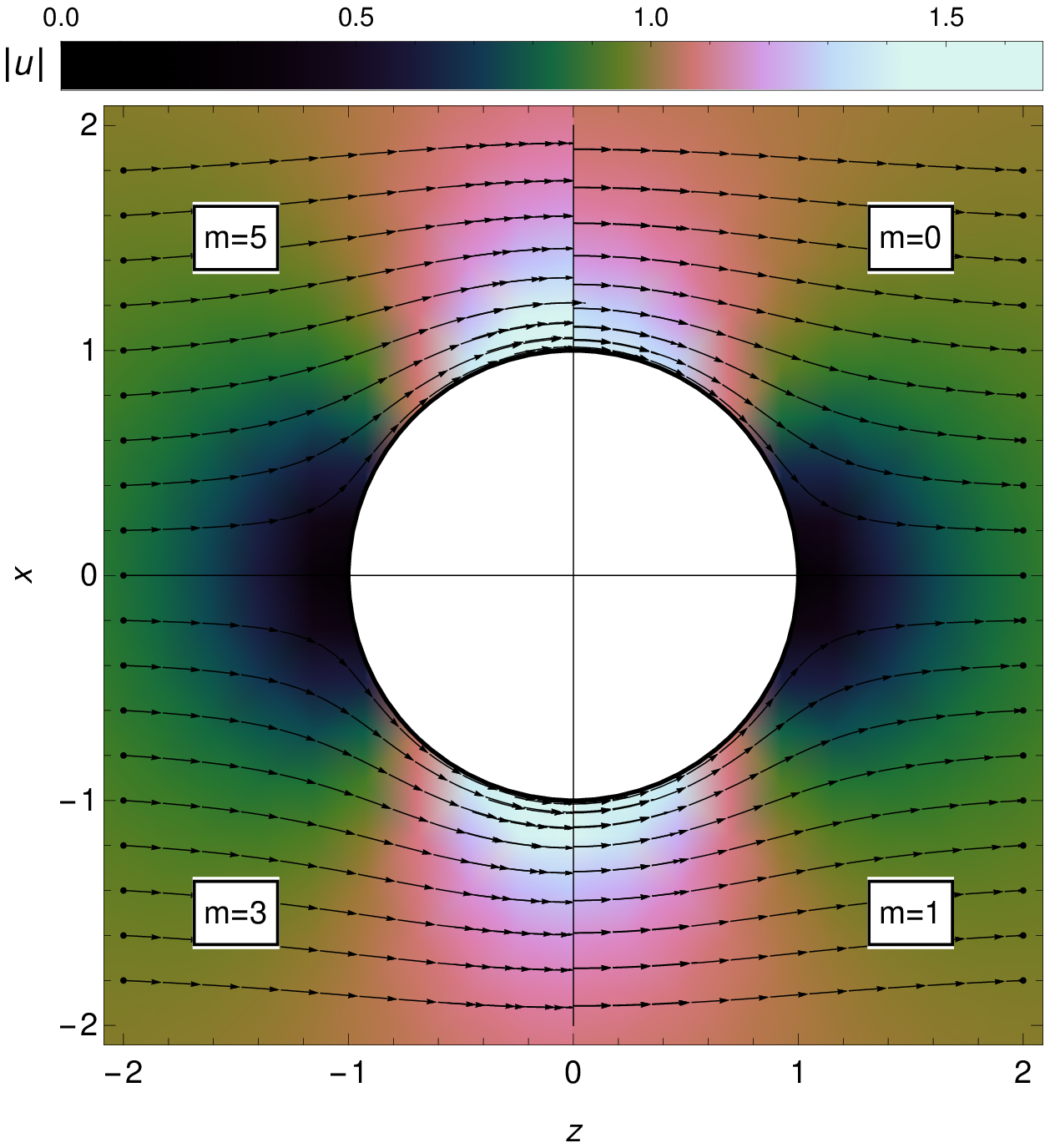}
\caption{
JRE solution to the critical flow around a unit-hypersphere (white disk) for $\gamma=7/5$ in 2D (a unit disk; left panel; incident near-critical Mach number $\mach_\infty=\mach^{\rm{disk}}_c \simeq 0.3982$) and 3D (a unit sphere; right panel; $\mach^{\rm{sphere}}_c \simeq 0.5619$).
In such a critical flow, the Mach number at the equator of the body locally reaches $\mach_{\rm{eq}}=1$.
Streamlines (arrows) represent the trajectory of the flow (passing through equidistant points at $z=\pm2$ (black dots).
The normalized (dimensionless) speed $|u|$ is computed up to different JRE orders $m$ (of $\mach_\infty^2$ in (\ref{eq: JRE for flow potential}); see labels) in each quadrant.
The effect of compressibility is particularly noticeable by comparing the $m=0$ and $m=5$ approximations along the $x>0$ axis.
}
\label{fig:flow_disk_sphere}
\end{figure*}

Figure \ref{fig:flow_disk_sphere} shows the flow around a disk (left panel) and a sphere (right panel) for incident Mach numbers at infinity approaching the respective critical/sonic Mach numbers.
The latter are tuned to yield a sonic flow at the equator of the hypersphere, $\mach(r=1,\theta=\pi/2)=1$, leading to $\mach^{\rm{disk}}_c \simeq 0.3982$ in 2D and $\mach^{\rm{sphere}}_c \simeq 0.5619$ in 3D.
The color intensity \citep[cubehelix;][]{Green11_Cubehelix} in the figure represents the normalized speed $|u|$.
The flow is symmetric under both $x$ and $z$ reflections, \ie $\theta \to -\theta$ and $\theta \to \pi-\theta$, so it is sufficient to plot only one quadrant of the $x-z$ plane.
We thus utilize all four quadrants to show the differences between the JRE obtained up to different, $m=0$ to $m=5$, orders (see labels).
To show the JRE corrections to the flow, we plot the streamlines (arrows) that pass through a set of equidistant points at $x=\pm2$, so the differences between quadrant arrows are meaningful.
Comparing the different quadrants shows that the compressible effects captured by higher orders $m$ raise the velocity and lower the density in the equatorial plane; most but not all of the change already transpires as $m=0$ is raised to $m=1$.

\begin{figure*}
  \includegraphics[width=\linewidth]{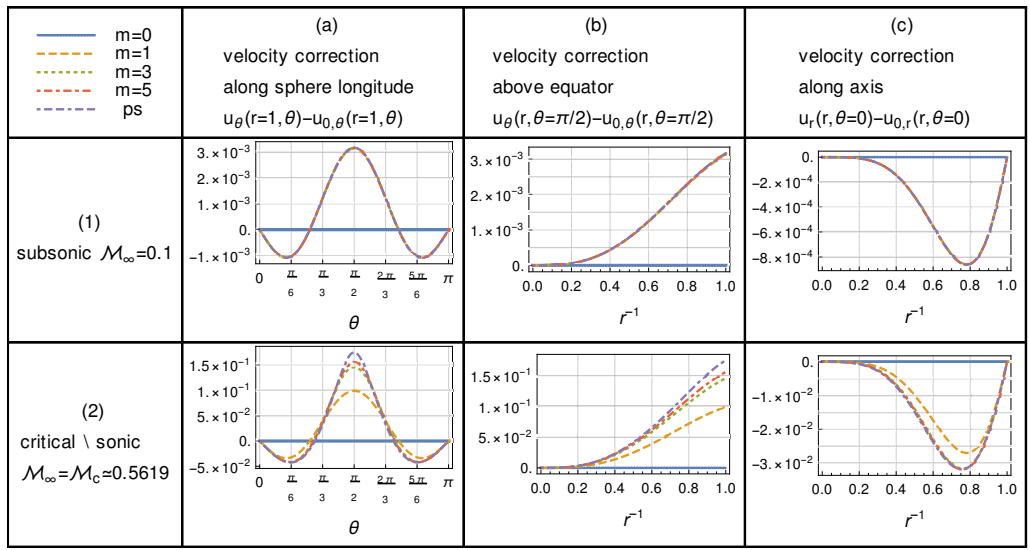}
\caption{
Compressible corrections to the flow around a 3D sphere for $\gamma=7/5$ according to different JRE orders and to the pseudospectral (ps) solver (see legend).
Profiles are shown for the flow along the longitude of the sphere (column a), at different radii above the equator (column b), and along the symmetry axis (column c), for both a subsonic flow with $\mach_\infty=0.1$ (row 1) and for the critical flow (row 2).
}
\label{fig:spherecompare}
\end{figure*}

Figure \ref{fig:spherecompare}
shows the compressible contribution to the flow around a sphere; qualitatively similar results are obtained for the disk.
The contribution is computed based on JREs of different orders, as well as on the numerical solution, and shown for the polar velocity on the sphere (column a), the polar velocity at the equatorial plane (column b), and the radial velocity along the symmetry axis (column c).
We plot these profiles for two different flows, with incident Mach numbers $\mach_\infty=0.1$ to demonstrate a subsonic case (row 1), and $\mach_\infty=\mach_c\simeq0.5619$ for the critical case (row 2).
As seen from row 1, at low Mach numbers, the JRE converges rapidly; the $m=1$ JRE is sufficient for accurately (within $\sim 0.005\%$ in $\bm{u}$ for $\mach_\infty=0.1$) capturing the compressible effects.
Convergence is slower for larger $\mach_\infty$, but manageable (compressible effects captured within $\sim 5\%$ in $\bm{u}$ for $m=1$) even in the critical limit.

For the flow around a disk in 2D, we find no logarithmic terms in the flow potential, for any $\gamma$.
Namely, all computed JRE (\ref{eq: JRE for potential 2D around disk}) coefficients $a$ with $\ell\neq 0$ vanish, as illustrated in table \ref{table: JRE coefficients for 2D around disk} up to $m=3$ order.
We confirm this behavior for arbitrary $\gamma$ up to order $m=30$.
Using a prescribed $\gamma$ speeds up the JRE computations, allowing us to reach higher orders.
We thus compute the JRE up to order $m=50$ for the specific cases $\gamma \in \{1, 7/5, 5/3, 2\}$, corresponding respectively to isothermal, ideal diatomic, ideal monatomic, and weakly-interacting bose, gasses.
In all of these cases, we find no logarithmic terms in the flow potential.

In contrast, logarithmic terms are unavoidable in the flow around a sphere in 3D, for any $\gamma$.
Indeed, the JRE (\ref{eq: JRE for potential 3D around sphere}) shows the first logarithmic term at order $m=3$, for $k=-8$, $n=7$ and $\ell=1$, as indicated in table \ref{table: JRE coefficients for 3D around sphere 2}.
This coefficient is proportional to $5+7\gamma+2\gamma^2$, which vanishes only for negative, non-physical $\gamma$ values.
In addition to such $\ell=1$ terms for order $m\geq3$, we find $\ell=2$ terms for order $m\geq 6$, $\ell=3$ terms for order $m\geq9$, and so on, with the highest logarithmic term order increasing by one every three orders in $m$.
Furthermore, the term with the highest logarithm power is also proportional to $\propto{r^{-2m-2}P_{2m+1}(\mu)}$.
This behavior is verified up to order $m=18$ for arbitrary $\gamma>-1$, and up to order $m=30$ for the specific values of $\gamma$ chosen above.

While proving that these effects persist as $m$ increases to infinity is beyond the scope of the present work, we hypothesize that they do:

\begin{conj}
\label{conj: disk no logarithm}
Logarithmic JRE terms never appear in the flow of a polytropic $\gamma \geq 1$ fluid around a disk.
\end{conj}

\begin{conj}
\label{conj: sphere logarithm appearance}
For the flow around a sphere, the highest $\ell$ of JRE terms $\propto{\mach_\infty^{2m}}{{\ln }^\ell(r)}$ is $\ell_{max}=\lfloor m/3 \rfloor$.
\end{conj}

\section{Example: axial hodographic approximation}
\label{sec:Application example: axial hodographic approximation}

\begin{figure*}
\subfloat[\label{fig:JRE Hodogrpahic comparison gamma 7/5}]{
  \includegraphics[width=.49\linewidth]{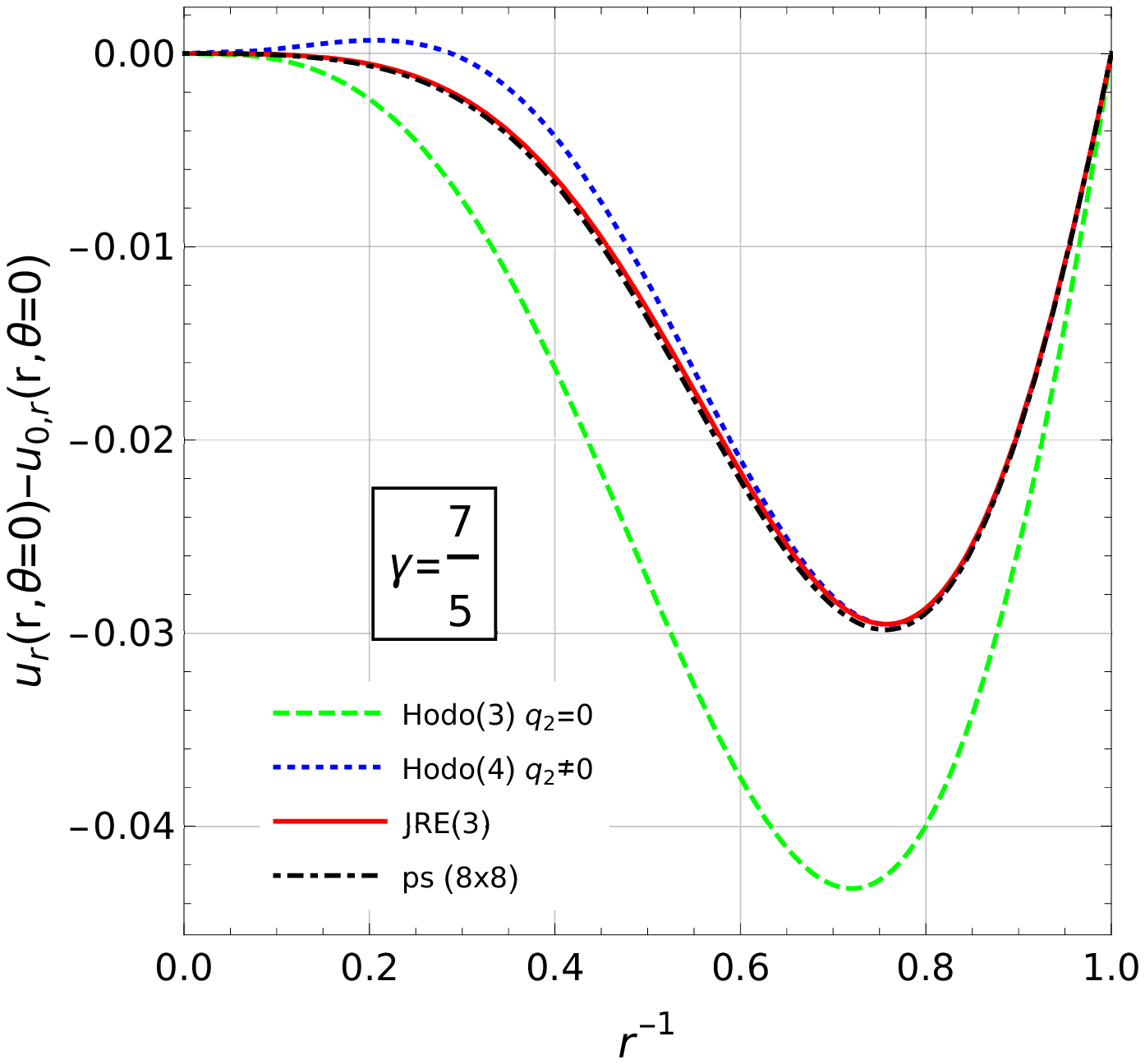}
  }\hfill
\subfloat[\label{fig:JRE Hodogrpahic comparison gamma 5/3}]{
  \includegraphics[width=.49\linewidth]{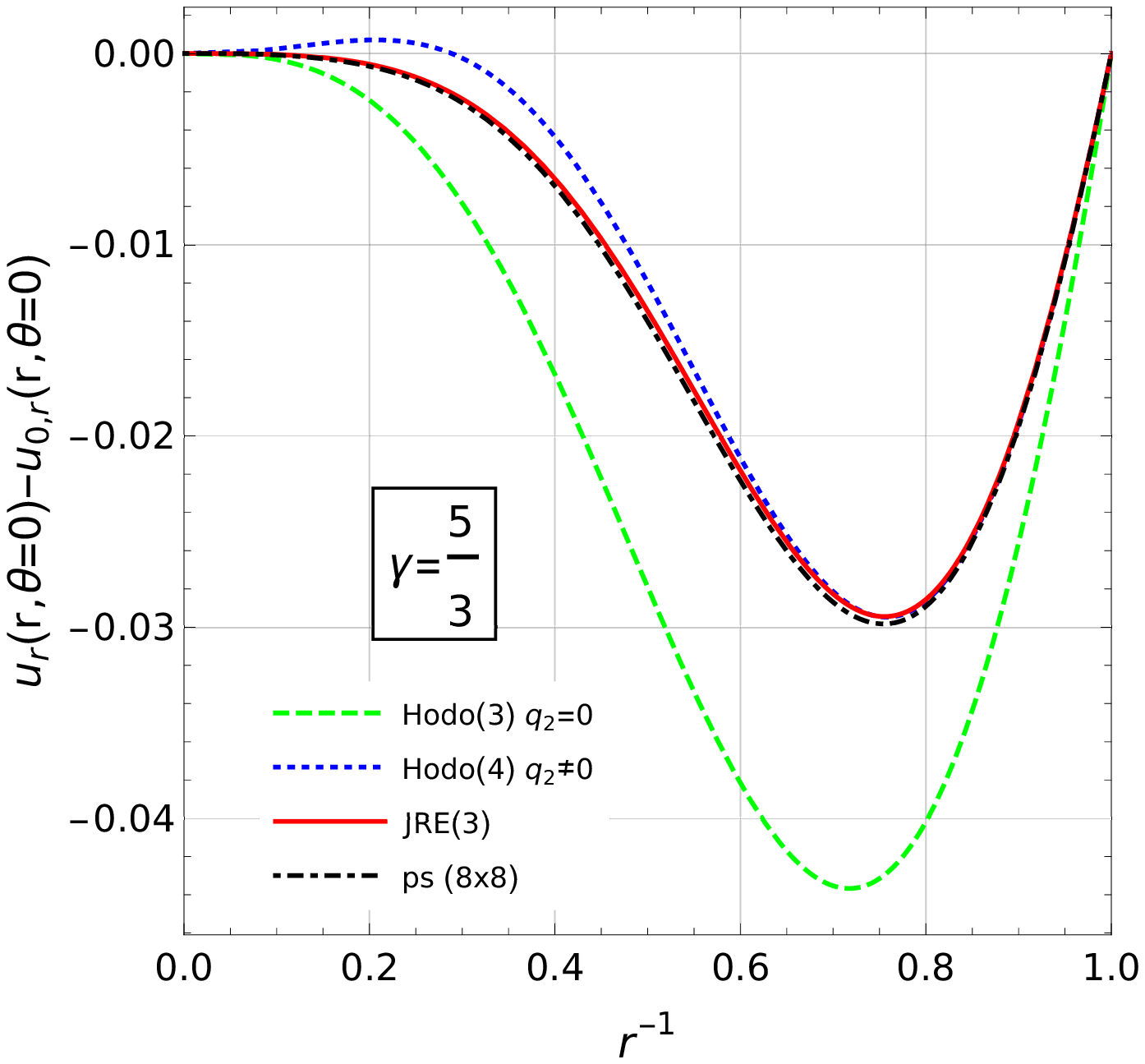}
}
\caption{
    Compressibility contribution to the velocity along the symmetry axis of a sphere near the critical Mach number, for $\gamma=7/5$ (left panel, $\mach_\infty=\mach_c\simeq0.5619$) and $\gamma=5/3$ (right panel, $\mach_\infty=\mach_c\simeq0.5462$).
    Results shown for a converged pseudospectral calculation (ps, dotted-dashes black curve), a third order JRE (JRE(3), solid red) and hodographic approximations with (Hodo(4), dotted blue) and without (Hodo(3), dashed green) the $q_2$ coefficient.
    \label{fig:JRE Hodogrpahic comparison gamma_Both}
    }
\end{figure*}

\begin{table}
  \begin{center}
  \def~{\hphantom{0}}
  \begin{tabular}{lccccc}
      & $m=0$ & $m=1$ & $m=2$ & $m=3$ & $m=4$ \\
      \hline
      $q_0$ & $1.5$ & $1.3809$ & $1.3783$ & $1.3766$ & $1.3762$\\
      $q_1$ & $-0.5$ & $-0.5$ & $-0.5$ & $-0.5$ & $-0.5$\\
      $q_2$ & $0$ & $0.0254$ & $0.0259$ & $0.0263$& $0.0265$\\
      $q_3$ & $0$ & $0.0585$ & $0.0662$ & $0.0689$& $0.0688$\\
  \end{tabular}
    \caption{Coefficients of the first four terms in the Taylor expansion of $q\left(u\right)$, for different JRE orders $m$, at the critical Mach number $\mach_\infty=\mach_{\infty}\simeq 0.5619$ for the flow of a $\gamma=7/5$ fluid around a sphere.
    See \S\ref{sec:Hodographic approximation of the solution of radialflow} for analytic expressions for $q$ with arbitrary $\mach_\infty$ and $\gamma$.}
    \label{table:qinumeric}
  \end{center}
 \end{table}

The flow in front of a blunt body has important implications in diverse applications, and is well approximated in both subsonic and supersonic regimes by expanding the perpendicular gradients $q\equiv\partial_\theta u_\theta|_{\theta=0}$ in terms of the parallel (normalized) velocity $u=|u_r|$  \citep{Keshet_Naor}
In the subsonic regime of this hodographic approximation, $q=q(u)$ is defined in the $0<u<1$ region between stagnation and spatial infinity, and is expanded around stagnation in the form $q\left(u\right) \approx q_0+q_1u+q_2u^2+q_3u^3$, which we designate as a 3rd order hodographic approximation, Hodo(3).
It can be shown that $q_0$ does not vary much with respect to the incompressible case, $q_1=-1/2$, and $q_2$ is small with respect to $q_3$ \citep{Keshet_Naor}.

Using the JRE for the sphere, here we calculate the coefficients $q_i$ analytically, for given order $m$.
The resulting expressions for the coefficients are provided for $1\leq m\leq 4$, as a function $\mach_\infty$ and $\gamma$, in appendix \S\ref{sec:Hodographic approximation of the solution of radialflow}.
For illustration, table \ref{table:qinumeric} provides the numerical values of these coefficients for the critical Mach number with $\gamma=7/5$.
The coefficients converge rapidly, reaching three-digit accuracy for $m=4$.
The effect of compressibility can be seen to be small, as $q_2$ and $q_3$ are smaller than $q_0$ and $q_1$ by two orders of magnitude.
We confirm the small deviation of $q_0$ from its incompressible value $3/2$, the precise result $q_1=-1/2$ for all orders, and that $q_2\approx q_3/3$ is small albeit non-negligible.

Figure \ref{fig:JRE Hodogrpahic comparison gamma_Both} shows the compressible contribution to the radial velocity along the symmetry axis of a sphere, with and without the JRE corrections, for a flow at the critical Mach number.
Results are shown both for $\gamma=7/5$ (left panel) and $\gamma=5/3$ (right panel), based on the pseudospectral code at resolution of $(8,8)$ (dotted-dashed curve), the full JRE of order $m=3$ (solid), on the hodographic approximation of \citet[][dashed]{Keshet_Naor}, and on our improved hodographic approximation (dotted).
The corrected hodographic approximation provides a much better fit to the JRE and to the actual flow.
The two panels slightly differ because they pertain to different $\mach_\infty$ values; for the same $\mach_\infty$, the axial flow for different $1<\gamma<2$ values are nearly indistinguishable.

\section{Example: flow in front of spheroids}
\label{sec:Flow around spheroids}

\begin{figure}
	\centering
	\includegraphics[width=.86\linewidth]{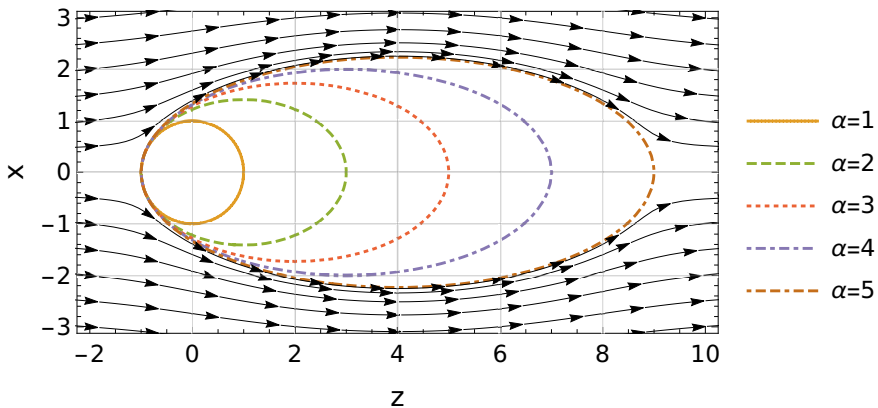}
	\caption{
    Different spheroids of the form $\left(x^2+y^2\right)/\alpha+\left(z-1+\alpha\right)^2/\alpha^2=1$ in the $y=0$ plane, along with streamlines (black arrows) of an incompressible flow around the most prolate body.
    The bodies are shifted such that their nose overlaps with the unit sphere.
    }
	\label{fig:SpheroidShapes}
\end{figure}

It has been suggested \citep{Keshet_Naor} that the flow in front of an axisymmetric body is well approximated by the flow in front of a sphere, rescaled to give the same nose curvature.
Consider general prolate or oblate spheroids of the form $\left(x^2+y^2\right)/\alpha+z^2/\alpha^2=1$, chosen to be axisymmetric along the $\hat{z}$, flow axis, and with unit curvature at the nose.
We shift the $z$ coordinate such that the resulting spheroid overlaps with the unit sphere at the nose $z \to z+1-\alpha^2$.
The pseudospectral code (details in \S\ref{sec:details and convergence of the pseudospectral solver}) is modified to solve the flow around such spheroids.

\begin{figure*}
\subfloat[\label{fig:spheroidvelincom}]{
  \includegraphics[width=.49\linewidth]{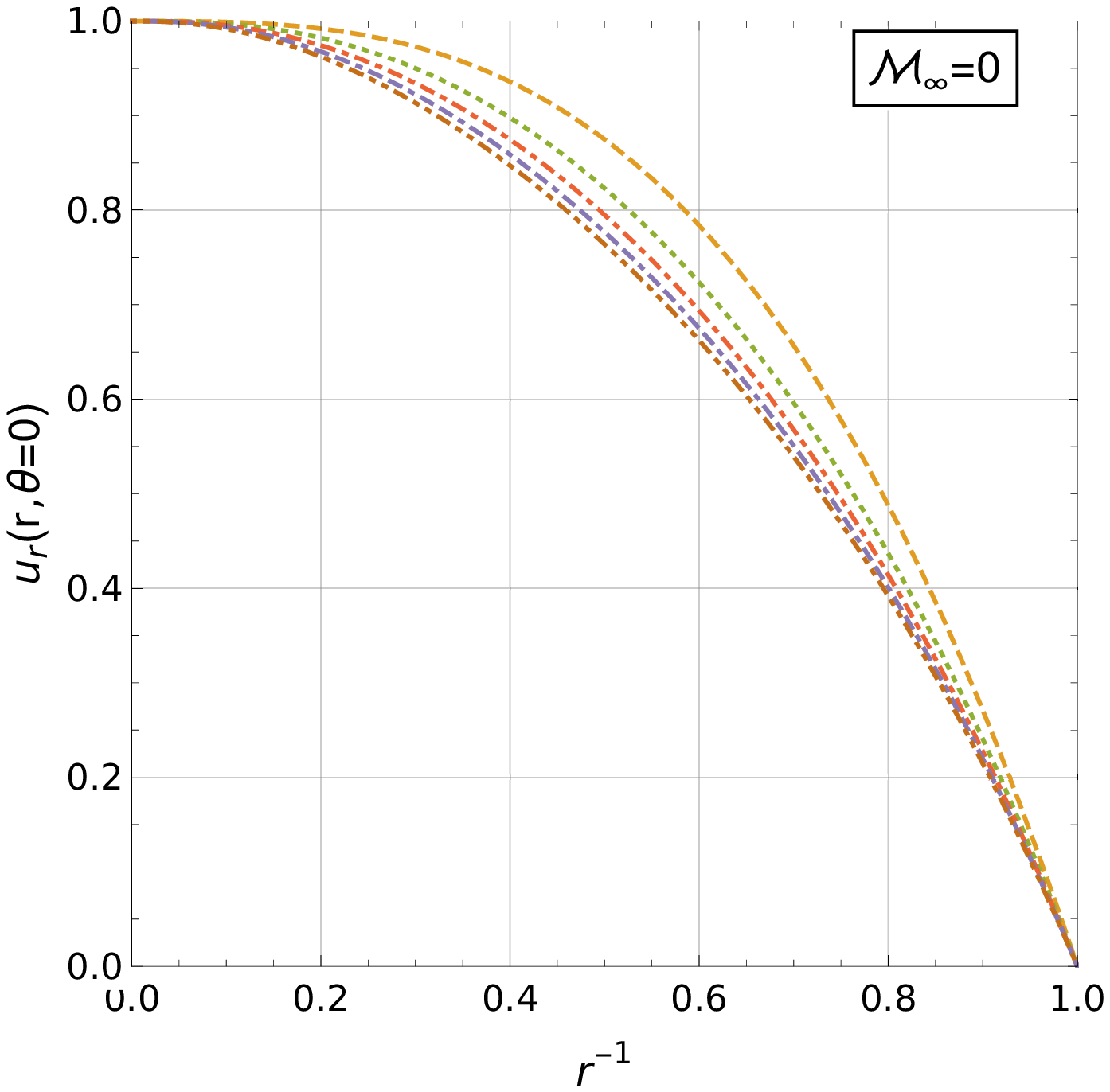}
}\hfill
\subfloat[\label{fig:spheroidvelsonic}]{
  \includegraphics[width=.49\linewidth]{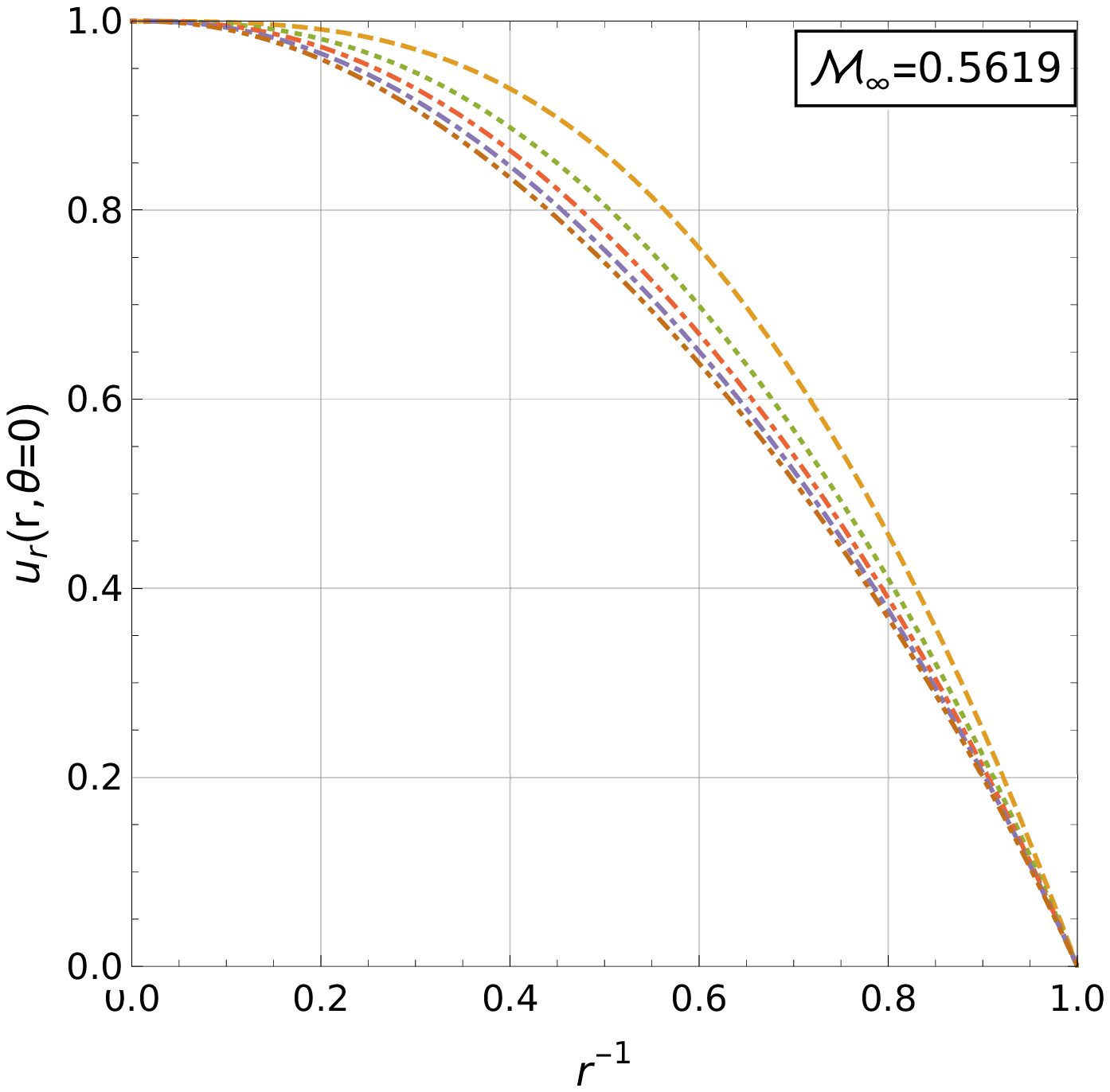}
}\hfill
\subfloat[\label{fig:spheroidvelincomscaled}]{
  \includegraphics[width=.49\linewidth]{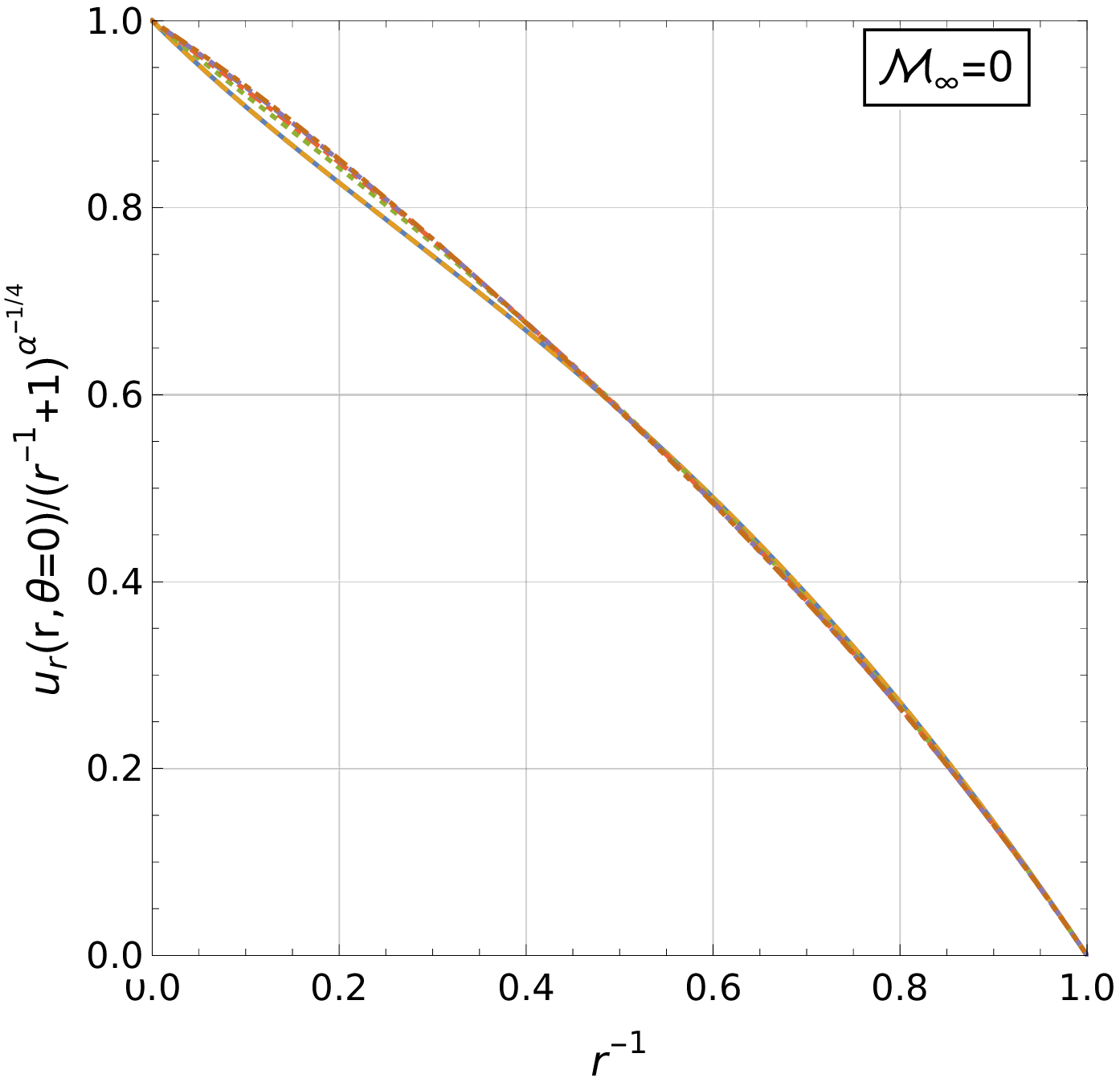}
}\hfill
\subfloat[\label{fig:spheroidvelsonicscaled}]{
  \includegraphics[width=.49\linewidth]{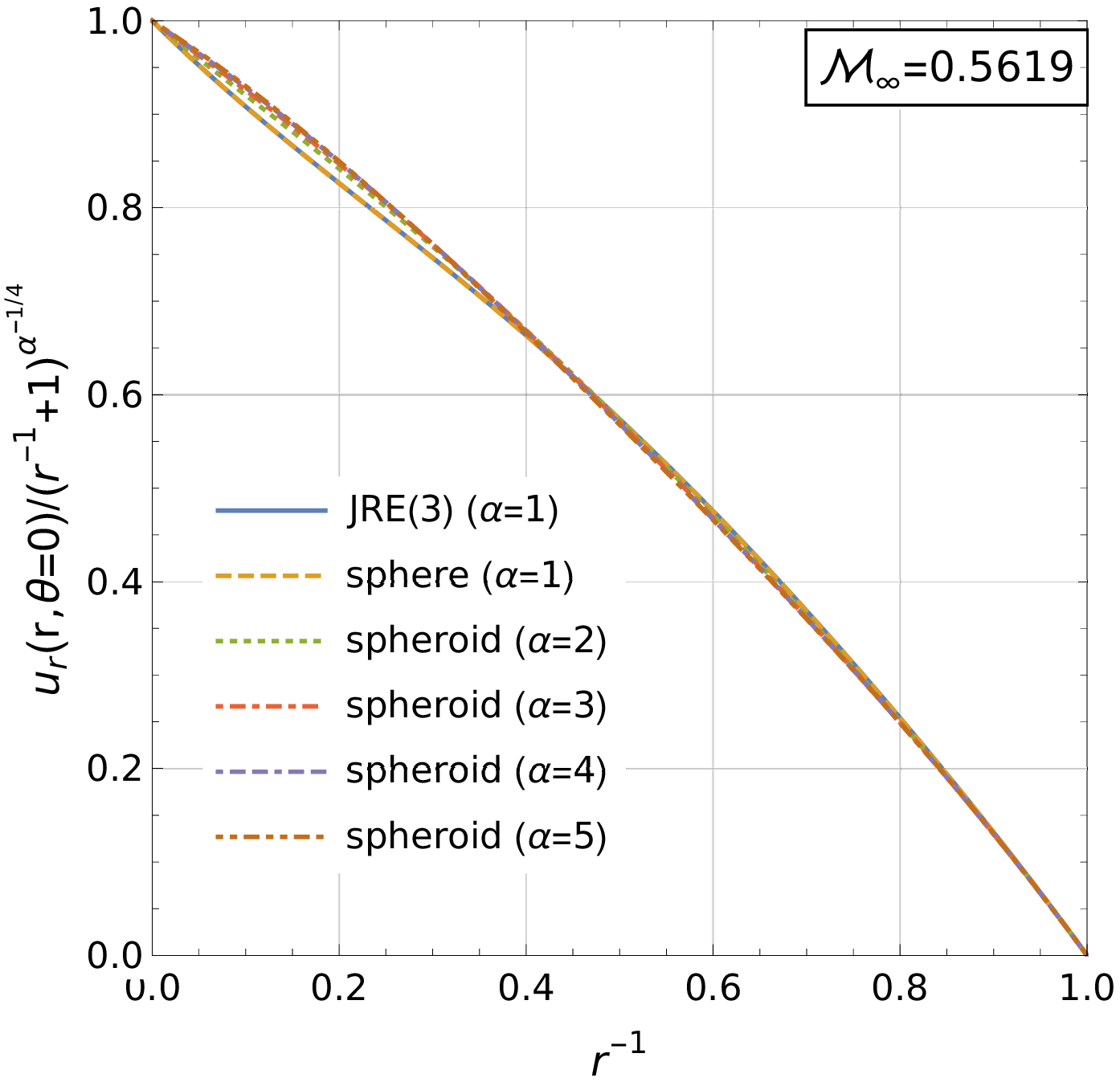}
}
\caption{
\label{fig:spheroidvelAll}
Incompressible (left panels) and critical (for the sphere: $\mach_\infty=0.5619$; right panels) flows in front of the different spheroids of figure \ref{fig:SpheroidShapes} (legend), showing the normalized velocity $u$ (upper panels) and the scaled velocity $u_r/\left(r^{-1}+1\right)^{\alpha^{-1/4}}$ (bottom panels).
Results are based on the pseudospectral code with a converged, $\left(k_{\max},n_{\max}\right)=\left(8,8\right)$ resolution, for $\gamma=7/5$.
In the bottom row we plot the third-order JRE around a sphere for comparison (JRE(3), solid blue).
}
\end{figure*}

Figure \ref{fig:SpheroidShapes} demonstrates the shapes of a sphere ($\alpha=1$) and of four shifted prolate spheroids ($\alpha \in \{2,3,4,5\}$), along with the incompressible flow along the latter, most prolate case.
Figure \ref{fig:spheroidvelAll} shows both the incompressible (left panels) and the critical (right panels) flows in front of these bodies.
The normalized velocity $u$ (upper panels) varies with $\alpha$, $\mach_\infty$ (and slightly with $\gamma$, see figure \ref{fig:spheroidvelsonicscaled} and \ref{fig:JRE Hodogrpahic comparison gamma_Both}).
However, a nearly universal result is obtained for the scaled velocity
\begin{equation}
\label{eq:flow in front spheroids scaling}
    u_r^{\rm{universal}}(r) \equiv u_r^{(\alpha)}\left(r,\theta=0\right)/\left(r^{-1}+1\right)^{\alpha^{-1/4}}\,,
\end{equation}
approximately insensitive to the Mach number, prolate body profile, and value of $\gamma$.
For oblate spheroids ($\alpha<1$), the velocity does not scale to the universal curve.

\section{Summary and discussion}
\label{sec:Summary and discussion}

We generalize the Janzen-Rayleigh expansion (JRE) of a hypersphere to arbitrary dimension, providing an exhaustive solution that supplements previous approaches with additional, usually necessary, logarithmic radial terms.
Such a generalization is found to be essential in 3D, required for obtaining the correct solution for the sphere even at third ($m=3$, \ie $\mach_\infty^6$) order, although it is apparently not needed for the special case of a disk in 2D.
The resulting, arbitrary-order JRE is a useful tool for studying various problems, as we demonstrate by generalizing and extending previous solutions for the axial flow of a sphere, and by presenting a simple approximate scaling that generalizes the axial flow for prolate spheroids.

The JRE of a potential, compressible flow is derived in general (in \S\ref{sec:Janzen-Rayleigh Expansion}) and around a hypersphere of arbitrary dimension $d$ (in \S\ref{sec:JRE for a hypersphere}).
The expansion is based on combining the continuity equation (\ref{eq: modified continuity}) and the Bernoulli equation (\ref{eq: Bernoulli principle}) into a single equation (\ref{eq: normalized flow potential non-linear PDE}) for the flow potential, $\phi$, which depends on the incident  Mach number $\mach_\infty$ far away from the hypersphere.
An expansion of $\phi$ in powers of the $\mach_\infty$ results in a set of recursive equations (\ref{eq: JRE flow potential linear PDE}), with the zero order being the incompressible solution (\ref{eq: flow potential m=0 solution}).
The equations for higher orders, $\phi_m$, are derived as a sum of particular Poisson equations (\ref{eq: JRE PDE genral Poisson source with logs}), with an exhaustive  solution that is a finite sum (\ref{eq: JRE PDE m>2 total solution}) of terms combining powers of $r$, Jacobi polynomials $\mj^{(d)}_n(\cos\theta)$,  and, in addition, powers of $\ln{(r)}$ that emerge when the Poisson source terms satisfy one of the special conditions (\ref{eq:BadK}).

Previous JRE calculations were typically carried out with only part of the general solution, each order being considered as a sum of powers of $r$ and Jacobi polynomials alone.
This approach works well in 2D, where one does not encounter any divergence when thus avoiding logarithmic terms, at least up to order $m=50$.
However, this approach severely limits the JRE in 3D, as without logarithmic terms, the $m=3$ order diverges.
The inclusion of logarithmic terms allows us to calculate terms much higher than available for any previous work in 3D \citep{kaplan1940flow, Tamada, Lighthill_1960,Fuhs_1976}, and with higher accuracy \citep{Frolov_2003}.
Furthermore, we are able to compute the JRE to arbitrary order in any dimension, providing interesting insights on the flow.
For instance, the JRE of the 4D hypersphere shows logarithmic terms at order $m=2$, indicating that the 3D sphere is not unique in requiring such terms.

After deriving the general solutions to all possible Poisson equations that can emerge in the problem (\ref{eq: JRE PDE inhomogeneous solution with logs}, \ref{eq: JRE PDE inhomogeneous solution with logs special}, and \ref{eq: JRE PDE inhomogeneous solution with logs special B}), we calculate the JRE analytically for general $\gamma$ to order $m=30$ in 2D and $m=18$ in 3D.
For select values of $\gamma$, we proceed to compute $50$ JRE orders in 2D and $30$ orders in 3D.

The $m=3$ order JRE includes a $\gamma$-dependent logarithmic term  proportional to $5+7\gamma+2\gamma^2$ (table \ref{table: JRE coefficients for 3D around sphere 2}), which does not vanish for $\gamma=1$ (\ie $w=0$).
This shows that both non-linear terms on the RHS of (\ref{eq: JRE flow potential linear PDE}) contribute to the logarithm.
We find that the highest order of any logarithmic term in the expansion increases by one every three orders in $m$,
and when this occurs, only one logarithmic term appears, multiplied by a single power of $r$ and a single Legendre polynomial $\propto{r^{-2m-2}P_{2m+1}(\mu)}$ (see table \ref{table: JRE coefficients for 3D around sphere 2} for $m=3$).
We speculate that these findings are true for higher orders in the JRE and all physical values of $\gamma$ (conjecture \ref{conj: sphere logarithm appearance}).

We expect the absence of logarithmic terms in the 2D disk JRE to persist for all orders $m$ and for all $\gamma$ (conjecture \ref{conj: disk no logarithm}).
The absence of logarithmic terms is the disk JRE is not, however, a property of 2D flows in general.
For instance, consider the incompressible flow of potential $\phi_0 = (r+1/r)\cos\theta+B_3r^{-3}\cos(3\theta)$ around an algebraic body, which is defined by the no-slip BCs and this potential, unique in leading to a finite polynomial equation (Wallerstein \& Keshet, in preparation). Here, $B_3$ is a coefficient in the general solution of the Laplace equation (\ref{eq: laplace general solution}).
The non-circular shape of the body leads to an infinite number of homogeneous terms in the first-order JRE potential $\phi_1$, and thus to an infinite number of both inhomogeneous and homogeneous terms in $\phi_2$.
It is nevertheless possible to isolate the finite number of terms that contribute to the nonzero term $\propto{r^{-5}\cos(5\theta)\ln(r)}$ in $\phi_2$.

The absence of logarithmic terms in the disk JRE thus appears to indicate an additional symmetry unique to this 2D body.
To illustrate this symmetry, consider how logarithmic terms could have putatively emerged in this JRE.
The $m=0$ potential $\phi_0=(r+r^{-1})\cos\theta$ induces the $m=1$ Poisson source $2(r^{-7}-2r^{-5})\cos\theta+2r^{-3}\cos 3\theta$ in (\ref{eq: JRE flow potential linear PDE}), none of which components satisfies (\ref{eq:BadK}), so $\phi_1$ has no logarithms.
This source curiously lacks terms such as $r^{-5}\cos3\theta$, which would have produced a logarithm in $\phi_1$.
Such a ($k=-5$, $n=3$) term is conspicuously absent from the ($m=2$, $n=3$) Poisson source $[3r^{-11}-2(2+3\gamma)r^{-9}+18\gamma r^{-7}+19r^{-3}]\cos(3\theta)/6$.
Inspecting (\ref{eq: JRE flow potential linear PDE}), one sees several combinations $(m_i,k_i,n_i)$ of terms among $\phi_{{m_1}}$, $\phi_{{m_2}}$, and $\phi_{{m_3}}$, each of which produces, in the $\phi_2$ source, a term
\begin{equation}
{\scriptsize f\begin{pmatrix} m_1 & k_1 & n_1 \\ m_2 & k_2 & n_2 \\ m_3 & k_3 & n_3 \end{pmatrix}} r^{-5}\cos3\theta \,,
\end{equation}
where $f$ are numerical coefficients.
However, the sum of these coefficients vanishes,
\begin{eqnarray}
& {\scriptsize f \begin{pmatrix} 1 & -3 & 1 \\ 0 & 1 & 1 \\ 0 & 1 & 1 \end{pmatrix}}+
{\scriptsize f \begin{pmatrix} 0 & -1 & 1 \\ 0 & 1 & 1 \\ 1 & -1 & 1 \end{pmatrix}}+
{\scriptsize f \begin{pmatrix} 0 & -1 & 1 \\ 1 & -3 & 3 \\ 0 & 1 & 1 \end{pmatrix}}+
{\scriptsize f \begin{pmatrix} 0 & 1 & 1 \\ 0 & 1 & 1 \\ 1 & -3 & 1 \end{pmatrix}} \\ & \quad\quad\quad +
{\scriptsize f \begin{pmatrix} 0 & 1 & 1 \\ 1 & -1 & 3 \\ 0 & 1 & 1 \end{pmatrix}} +
{\scriptsize f \begin{pmatrix} 0 & -1 & 1 \\ 0 & -1 & 1 \\ 1 & -1 & 3 \end{pmatrix}}
=\frac{3}{2}-\frac{3}{4}+\frac{3}{2}-\frac{3}{4}-\frac{3}{4}-\frac{3}{4}=0 \, , \nonumber
\end{eqnarray}
leading to the absence of logarithms in $\phi_2$. Such a cancellation of terms persists at higher orders $m$.

Regardless of logarithmic terms, the angular part the solution includes Jacobi polynomials of odd $n$ for all JRE orders, so the flow is symmetric fore and aft of the hypersphere.
Hence, as long as the JRE converges and no additional effects are included, there is no net force on the hypersphere, and the flow is drag-free for all dimensions.
The d'Alembert paradox thus persists in all dimensions.

Even though we calculate the JRE to high orders, many physical features can be adequately captured with only a few low orders (figure \ref{fig:spherecompare}), even for a sonic flows.
For example, we find an accuracy better than $\sim 1\%$ in $\bm{u}$ for $m=5$.
The low JRE orders ($m \leq 5$) show that compressibility effects are more prominent in the vicinity of the hypersphere, and in particular at the equatorial plane (figures \ref{fig:flow_disk_sphere} and \ref{fig:spherecompare}).

With the ability to calculate the JRE to any desired accuracy, we compare it to other approximation methods, such as the hodographic approximation for the flow on the symmetry axis in front of a sphere, which has important physical implications even in the inviscid regime \citep[\eg][]{Keshet_Naor}.
The hodographic approximation of \citet{Keshet_Naor} performs better than the JRE of order $m=0$, but worse than the JRE of order $m=1$ for any $\gamma$ (figure \ref{fig:JRE Hodogrpahic comparison gamma 7/5} and \ref{fig:JRE Hodogrpahic comparison gamma 5/3}).
We use the JRE to improve the hodographic approximation (appendix \S\ref{sec:Hodographic approximation of the solution of radialflow}), such that it performs better than JRE of order $m=2$, but worse than the JRE of order $m=3$ again for any $\gamma$, but can be continued to the supersonic regime \citep{Keshet_Naor}.

It has been speculated that the flow in front of the sphere can be applied to axisymmetric bodies with a similarly scaled nose curvature  \citep[][and references therein]{Keshet_Naor},
such as prolate spheroids (figure \ref{fig:SpheroidShapes}).
While the flows in front of different spheroids are not identical (figure \ref{fig:spheroidvelincom} and \ref{fig:spheroidvelsonic}), they can be approximately mapped onto each other with a universal scaling, independent of $\gamma$ and $\mach_\infty$, for flows ranging from the incompressible (figure \ref{fig:spheroidvelincomscaled}) to the sonic (\ref{fig:spheroidvelsonicscaled}) regimes.
The velocity $u_r^{(\alpha)}$ in front of prolate spheroids with semi-axes $\alpha>0$ and $\alpha^{1/2}$ can be well approximated by a scaled JRE for a sphere
\begin{eqnarray}
    \label{eq: flow in front spheroids, scaled + JRE}
    u_r^{(\alpha)}\left(r,\theta=0\right)&\approx&\left(r^{-1}+1\right)^{\alpha^{-1/4}-1} u_r^{(1)}\left(r,\theta=0\right)\\
    &\approx&\left(r^{-1}+1\right)^{\alpha^{-1/4}-1}\sum\limits_{m=0}^{m_{max}} \mach_\infty^{2m}\partial_{r}\phi_m\left(r,\theta=0\right). \nonumber
\end{eqnarray}
It is natural to ask if such scalings can be identified using the generalized JRE in front of other bodies and in other dimensions, but this is beyond the scope of the present work.

\vspace{0.5cm}

We thank Y. Lyubarsky, D. Kogan and E. Grosfeld for helpful discussions.
This research has received funding from the GIF (Grant No. I-1362-303.7 / 2016), and was supported by the Ministry of Science, Technology \& Space, Israel, by the IAEC-UPBC joint research foundation (Grants No. 257/14 and 300/18), and by the Israel Science Foundation (Grant No. 1769/15).

\appendix

\section{Explicit JRE coefficients for low orders}
\label{sec:Explicit JRE coefficients for low orders}

\begin{table}
\centering
\rotatebox{90}{\begin{minipage}{\textheight}
\begin{tabular}{lrrrrrrrrrr}
\toprule
$a_{m,k,n}$ & \multicolumn{1}{c}{$m=0$} & \multicolumn{2}{c}{$m=1$} & \multicolumn{3}{c}{$m=2$} & \multicolumn{4}{c}{$m=3$}\\
\cmidrule(lr){2-2}\cmidrule(lr){3-4}\cmidrule(lr){5-7}\cmidrule(lr){8-11}
& $n=1$  & $n=1$ & $n=3$ & $n=1$ & $n=3$ & $n=5$ & $n=1$ & $n=3$ & $n=5$ & $n=7$\\
\midrule
$k=+1$ & $1$ & & & & & & & & & \\\hline
$k=-1$ & $1$ & $\frac{13}{12}$ & $-\frac{1}{4}$ & $\frac{17 \gamma }{60}+\frac{343}{240}$ & $-\frac{19}{48}$ & $\frac{1}{16}$ & $\frac{59 \gamma ^2}{420}+\frac{41 \gamma }{42}+\frac{2273}{1120}$ & $-\frac{17 \gamma }{240}-\frac{137}{240}$ & $\frac{25}{192}$ & $-\frac{1}{64}$ \\\hline
$k=-3$ & & $-\frac{1}{2}$ & $\frac{1}{12}$ & $-\frac{\gamma }{8}-\frac{29}{24}$ & $\frac{3}{20}-\frac{61 \gamma }{240}$ & $\frac{\gamma }{16}$ & $-\frac{163 \gamma }{240}-\frac{1759}{720}$ & $-\frac{29 \gamma ^2}{210}-\frac{2117 \gamma }{2100}+\frac{583}{14400}$ & $\frac{97 \gamma }{240}+\frac{157}{960}$ & $-\frac{\gamma }{16}-\frac{1}{32}$ \\\hline
$k=-5$ & & $\frac{1}{12}$ & & $\frac{\gamma }{12}+\frac{35}{48}$ & $\frac{3 \gamma }{16}$ & $-\frac{3 \gamma }{80}-\frac{1}{80}$ & $\frac{\gamma ^2}{48}+\frac{47 \gamma }{64}+\frac{3719}{1440}$ & $\frac{\gamma ^2}{16}+\frac{387 \gamma }{320}+\frac{209}{960}$ & $\frac{169 \gamma ^2}{1680}-\frac{11131 \gamma }{33600}-\frac{60911}{302400}$ & $-\frac{\gamma ^2}{48}+\frac{7 \gamma }{192}+\frac{5}{192}$ \\\hline
$k=-7$ & & & & $-\frac{\gamma }{16}-\frac{1}{4}$ & $-\frac{\gamma }{40}-\frac{1}{60}$ & & $-\frac{\gamma ^2}{16}-\frac{749 \gamma }{960}-\frac{5443}{2880}$ & $-\frac{\gamma ^2}{40}-\frac{1493 \gamma }{2400}-\frac{349}{1800}$ & $-\frac{\gamma ^2}{12}+\frac{\gamma }{16}+\frac{7}{96}$ & $\frac{5 \gamma ^2}{336}+\frac{\gamma }{1344}-\frac{1}{336}$ \\\hline
$k=-9$ & & & & $\frac{\gamma }{80}+\frac{1}{30}$ & $\frac{1}{144}$ & & $\frac{3 \gamma ^2}{80}+\frac{149 \gamma }{320}+\frac{2449}{2880}$ & $\frac{\gamma ^2}{24}+\frac{281 \gamma }{1440}+\frac{143}{1080}$ & $\frac{\gamma ^2}{112}+\frac{13 \gamma }{2240}-\frac{13}{840}$ & \\\hline
$k=-11$ & & & & & & & $-\frac{\gamma ^2}{60}-\frac{17 \gamma }{120}-\frac{61}{288}$ & $-\frac{\gamma ^2}{112}-\frac{83 \gamma }{2240}-\frac{5}{96}$ & $\frac{1}{1440}-\frac{\gamma }{240}$ & \\\hline
$k=-13$ & & & & & & & $\frac{\gamma ^2}{336}+\frac{113 \gamma }{6720}+\frac{227}{10080}$ & $\frac{\gamma }{240}+\frac{11}{1440}$ & $\frac{1}{1728}$ & \\
\bottomrule
\end{tabular}
\caption{Coefficients table for the JRE up to third-order ($\mach_\infty^6$) in 2D (disk)}
\label{table: JRE coefficients for 2D around disk}
\end{minipage}}
\end{table}	

\begin{table}
\centering
\rotatebox{90}{
\begin{minipage}{\textheight}
\begin{scriptsize}
\begin{tabular}{lrrrrrrr}
\toprule
$b_{m,k,n,0}$ & \multicolumn{1}{c}{$m=0$} & \multicolumn{2}{c}{$m=1$} & \multicolumn{3}{c}{$m=2$} & $m=3$ \\
\cmidrule(lr){2-2}\cmidrule(lr){3-4}\cmidrule(lr){5-7}\cmidrule(lr){8-8}
& $n=1$  & $n=1$ & $n=3$ & $n=1$ & $n=3$ & $n=5$ & $n=1$\\
\midrule
$k=+1$ & $1$ & & & & & \\\hline
$k=-2$ & $\frac{1}{2}$ & $\frac{1}{3}$ & $-\frac{3}{10}$ & $\frac{599\gamma}{16800}+\frac{78733}{271040}$ & $-\frac{53}{150}$ & $\frac{5}{42}$ & $\frac{131\gamma ^2}{23520}+\frac{7507037 \gamma }{100793000}+\frac{1160213239}{4192988800}$ \\\hline
$k=-4$ & & & $\frac{27}{55}$ & & $\frac{83133\gamma}{3665200}+\frac{13691721}{19059040}$ & $-\frac{6}{11}$ \\\hline
$k=-5$ & & $-\frac{1}{5}$ & $-\frac{3}{10}$ & $-\frac{3\gamma}{70}-\frac{149}{525}$ & $-\frac{\gamma}{10}-\frac{23}{75}$ & $\frac{\gamma}{7}+\frac{11}{21}$ & \\\hline
$k=-5$ & & & & & & & $-\frac{2519 \gamma }{21000}-\frac{6013321}{15246000}$ \\\hline
$k=-6$ & & & & & & $\frac{4805247}{12172160}-\frac{50517\gamma}{276640}$ & \\\hline
$k=-7$ & & & & $-\frac{243}{1925}$ & $-\frac{156}{275}$ & $-\frac{60}{77}$ & $-\frac{4346757 \gamma }{128282000}-\frac{457071093}{3335332000}$ \\\hline
$k=-8$ & & $\frac{1}{24}$ & $\frac{3}{176}$ & $\frac{\gamma}{24}+\frac{1381}{4620}$ & $\frac{15\gamma }{176}+\frac{373}{880}$ & $\frac{3\gamma}{52}+\frac{211}{728}$ & $\frac{\gamma ^2}{105}+\frac{50143 \gamma }{246400}+\frac{26742953}{48787200}$ \\\hline
$k=-9$ & & & & & & $-\frac{3888}{21175}$ \\\hline
$k=-10$ & & & & $\frac{5589}{84700}$ & $\frac{1137}{7150}$ & $\frac{87}{1540}$ & $\frac{573186051 \gamma }{5644408000}+\frac{87731975043}{146754608000}$ \\\hline
$k=-11$ & & & & $-\frac{13\gamma}{560}-\frac{131}{924}$ & $-\frac{57 \gamma }{1960}-\frac{1075}{6468}$ & $-\frac{9 \gamma}{1120}-\frac{1697}{36960}$ & $-\frac{17 \gamma ^2}{840}-\frac{179359 \gamma }{646800}-\frac{1289}{1764}$ \\\hline
$k=-12$ & & & & & & & $\frac{16281}{125125}$ \\\hline
$k=-13$ & & & & & & & $-\frac{81 \gamma }{1078}-\frac{355023}{770770}$ \\\hline
$k=-14$ & & & & $\frac{13\gamma}{2800}+\frac{6073}{369600}$ & $\frac{23 \gamma }{6800}+\frac{2957}{224400}$ & $\frac{5\gamma}{8512}+\frac{1285}{561792}$ & $\frac{43 \gamma ^2}{2800}+\frac{29644609 \gamma }{153938400}+\frac{46207961}{92363040}$ \\\hline
$k=-16$ & & & & & & & $\frac{33183 \gamma }{1832600}+\frac{7173927}{104824720}$ \\\hline
$k=-17$ & & & & & & & $-\frac{\gamma ^2}{175}-\frac{7908301 \gamma }{128282000}-\frac{3261710029}{25399836000}$ \\\hline
$k=-20$ & & & & & & & $\frac{41 \gamma ^2}{47040}+\frac{2943929 \gamma }{439824000}+\frac{2032133}{188496000}$ \\
\bottomrule
\end{tabular}
\end{scriptsize}
\caption{Coefficients table for the third-order JRE ($\mach_\infty^6$) in 3D (sphere) (For the third-order only $n=1$ terms are present, see table \ref{table: JRE coefficients for 3D around sphere 2} for the remaining terms)}
\label{table: JRE coefficients for 3D around sphere 1}
\end{minipage}}
\end{table}

\begin{table}
\centering
\rotatebox{90}{
\begin{minipage}{\textheight}
\begin{scriptsize}
\begin{tabular}{lrrrr}
\toprule
$b_{m,k,n,0}$ & \multicolumn{3}{c}{$m=3$}\\
\cmidrule(lr){2-4}
& $n=3$ & $n=5$ & $n=7$\\
\midrule
$k=-2$ & $-\frac{599 \gamma }{28000}-\frac{22815547}{60984000}$ & $\frac{164}{819}$ & $-\frac{35}{858}$ \\\hline
$k=-4$ & $-\frac{418827051 \gamma ^2}{29673459200}+\frac{6207054461763 \gamma }{73293444224000}+\frac{9236174319103493}{10480962524032000}$ & $-\frac{9237 \gamma }{366520}-\frac{157349}{146608}$ & $\frac{567}{1573}$ \\\hline
$k=-5$ & $-\frac{104327 \gamma }{462000}-\frac{9833963}{30492000}$ & $\frac{661 \gamma }{1365}+\frac{4471}{4095}$ & $-\frac{30 \gamma }{143}-\frac{70}{143}$ \\\hline
$k=-6$ & & $-\frac{445149 \gamma ^2}{95190095}-\frac{587573895081 \gamma }{1424043821200}+\frac{104968047599}{125190665600}$ & $\frac{151551 \gamma }{513760}-\frac{14415741}{22605440}$ \\\hline
$k=-7$ & $-\frac{8118171 \gamma }{50396500}-\frac{114716983}{100793000}$ & $-\frac{1179657 \gamma }{3335332}-\frac{55771}{46648}$ & $\frac{756 \gamma }{1573}+\frac{2916}{1573}$ \\\hline
$k=-8$ & $\frac{61 \gamma ^2}{2420}+\frac{27461211 \gamma }{70470400}+\frac{3347992607}{4651046400}$ & $\frac{34 \gamma ^2}{1183}+\frac{2901 \gamma }{47320}-\frac{4841}{141960}$ & $\frac{427 \gamma ^2}{41184}-\frac{167888961127633 \gamma }{166256455622400}-\frac{5482054296482347}{7315284047385600}$ \\\hline
$k=-9$ & $\frac{16839 \gamma }{304304}-\frac{1085826767}{3682078400}$ & $\frac{33678 \gamma }{112385}-\frac{4676251}{5439434}$ & $\frac{1060857 \gamma }{2825680}-\frac{1647800217}{1367629120}$ \\\hline
$k=-10$ & $\frac{1149664413 \gamma }{5241236000}+\frac{207885400509}{136272136000}$ & $\frac{466297149 \gamma }{1334132800}+\frac{4846089009}{2668265600}$ & $\frac{12555 \gamma }{53482}+\frac{131139}{106964}$ \\\hline
$k=-11$ & $-\frac{361 \gamma ^2}{10780}-\frac{3681043 \gamma }{8408400}-\frac{19239877}{18498480}$ & $-\frac{19 \gamma ^2}{560}-\frac{18280747 \gamma }{54654600}-\frac{24731363}{36436400}$ & $-\frac{\gamma ^2}{66}-\frac{16589 \gamma }{111540}-\frac{125003}{408980}$ \\\hline
$k=-12$ & $\frac{30359621107}{147283136000}-\frac{488331 \gamma }{12172160}$ & $\frac{36101964719}{118362083840}-\frac{18202959 \gamma }{244549760}$ & $\frac{44618931927}{415759252480}-\frac{20762487 \gamma }{859006720}$ \\\hline
$k=-13$ & $-\frac{147 \gamma }{968}-\frac{646183}{692120}$ & $-\frac{9861 \gamma }{70070}-\frac{635053}{770770}$ & $-\frac{2997 \gamma }{78650}-\frac{192537}{865150}$ \\\hline
$k=-14$ & $\frac{1803 \gamma ^2}{74800}+\frac{1337811833 \gamma }{4596160800}+\frac{3982665971}{5515392960}$ & $\frac{111 \gamma ^2}{6916}+\frac{1078190297 \gamma }{5815249440}+\frac{3055176545}{6978299328}$ & $\frac{57 \gamma ^2}{16016}+\frac{1873881 \gamma }{45805760}+\frac{8865797}{91611520}$ \\\hline
$k=-16$ & $\frac{31497 \gamma }{919600}+\frac{6545551}{52601120}$ & $\frac{26589 \gamma }{1401400}+\frac{207773}{2802800}$ & $\frac{3699 \gamma }{1157728}+\frac{317763}{25470016}$ \\\hline
$k=-17$ & $-\frac{17 \gamma ^2}{2200}-\frac{31358782537 \gamma }{398333936000}-\frac{195575670263}{1195001808000}$ & $-\frac{305 \gamma ^2}{88088}-\frac{21471961559 \gamma }{613434261440}-\frac{136186343437}{1840302784320}$ & $-\frac{7 \gamma ^2}{13728}-\frac{295271 \gamma }{57383040}-\frac{623911}{57383040}$ \\\hline
$k=-20$ & $\frac{553 \gamma ^2}{647680}+\frac{2385587089 \gamma }{345181056000}+\frac{130642399631}{11390974848000}$ & $\frac{281 \gamma ^2}{1019200}+\frac{501680027 \gamma }{217273056000}+\frac{8141798513}{2048574528000}$ & $\frac{5 \gamma ^2}{164736}+\frac{35015 \gamma }{137719296}+\frac{1325845}{3029824512}$ \\\bottomrule
$b_{m,k,n,1}$ & & & \\\hline
$k=-8$ & & & $-\frac{6 \gamma ^2}{143}-\frac{21 \gamma }{143}-\frac{15}{143}$ \\
\bottomrule
\end{tabular}
\end{scriptsize}
\caption{Coefficients table for the third-order JRE ($\mach_\infty^6$) in 3D (sphere) (only $n\in\{3, 5, 7\}$ terms). The last row indicates the single logarithmic term.}
\label{table: JRE coefficients for 3D around sphere 2}
\end{minipage}}
\end{table}

Expressions for the coefficients of the JRE up to order $\mach_\infty^6$, valid for any $\gamma$, are provided
in table \ref{table: JRE coefficients for 2D around disk} for a disk, and in tables \ref{table: JRE coefficients for 3D around sphere 1} and \ref{table: JRE coefficients for 3D around sphere 2} for a sphere.
Each table is broken to parts, each corresponding to a different $m$ index of $\mach_\infty^{2m}$.
Columns correspond to the $n$ appearing in $\cos\left(n\theta\right)$ or $P_n\left(\cos\theta\right)$ basis functions, and the rows correspond to $k$ of the $r^k$ terms.
All coefficients in these tables pertain to non-logarithmic terms, \ie to $\propto\ln^\ell \left(r\right)$ terms with $\ell=0$, except for the very last term in table \ref{table: JRE coefficients for 3D around sphere 2}, which has $\ell=1$.

\section{Hodographic approximation of the solution of a radial flow}
\label{sec:Hodographic approximation of the solution of radialflow}

In \S\ref{sec:Application example: axial hodographic approximation} we discuss the hodographic approximation for the flow in front of a sphere, and use the JRE to obtain a more accurate expansion.
To relate between the radial velocity and the radius we use Eq. (7) from \citet{Keshet_Naor},
\begin{equation}
    \label{eq: hodographic apprximation for radial velocity}
    2 \ln\left(r\right)={\int }_{0}^{-u_r(r)} \frac{1-{\mach}_{0}{(u^{\prime} )}^{2}/{W}^{2}}{1-{\mach}_{0}{(u^{\prime} )}^{2}/{S}^{2}} \frac{{du}^{\prime} }{q(u^{\prime} )-u^{\prime} },
\end{equation}
with $W^2=2/\left(\gamma+1\right), S^2=2/\left(\gamma-1\right)$  (notice that $S^2=w^{-1}$) and
\[
    \mach_0\left(u\right)=u/\overline{c}=u/\left(S^{-2}+\mach_\infty^{-2}\right)^{1/2},
\]
which is the Mach number with respect to the stagnation point.
To complete the integral we calculate the coefficients $q_i$ for $i\in\{0,1,2,3\}$ using the JRE.

We provide explicit expressions for general $\gamma$ in table \ref{table: general hodographic coefficeints} which are valid for $0\leq\mach_\infty\leq\mach_c$ (with $q_1\equiv1/2$ identically \citep{Keshet_Naor}).
These values give a good approximation for the radial velocity near the stagnation point, but deviates far from the sphere.
To ensure the correct BC far from the body we add a fifth term to $q\left(u\right)$ such that the denominator in the integral in (\ref{eq: hodographic apprximation for radial velocity}) vanishes at $r\to\infty$ \ie $q_4=1-q_0-q_1-q_2-q_3$.

\begin{table}
\centering
\rotatebox{90}{
\begin{minipage}{\textheight}
\begin{scriptsize}
\begin{tabular}{lrrr}
\toprule
& $m=0$ & $m=1$ & $m=2$\\
\midrule
$q_0$ & $\frac{3}{2}$ & $\frac{3}{2}-\frac{83 \mach_\infty^2}{220}$ & $\frac{3}{2}-\frac{83 \mach_\infty^2}{220}-\frac{(378591917 \gamma -678984653) \mach_\infty^4}{5690484800}$ \\\hline
$q_2$ & $0$ & $\frac{-1408394988000 \mach_\infty^2 \left[\mach_\infty^2 (102855 \gamma -333287)-3527160\right]}{ \left[(378591917 \gamma -678984653) \mach_\infty^4-2146864720 \mach_\infty^2+8535727200\right]^2}$ & $\frac{-1408394988000 \mach_\infty^2 \left[\mach_\infty^2 (102855 \gamma -333287)-3527160\right]}{ \left[(378591917 \gamma -678984653) \mach_\infty^4-2146864720 \mach_\infty^2+8535727200\right]^2}$ \\\hline
$q_3$ & $0$ & $-\frac{7623000 \mach_\infty^2}{\left(83 \mach_\infty^2-330\right)^3}$ & $\frac{-8014450271610182400000 \mach_\infty^2 \left[\mach_\infty^2 (6569315 \gamma +2958729)-16460080\right]}{ \left[(378591917 \gamma -678984653) \mach_\infty^4-2146864720 \mach_\infty^2+8535727200\right]^3}$ \\\midrule
\end{tabular}

\begin{tabular}{lr}
& $m=3$ \\\midrule
$q_0$ & $\frac{3}{2}-\frac{83 \mach_\infty^2}{220}+\frac{(378591917 \gamma -678984653) \mach_\infty^4}{5690484800} -\frac{\left(1058753493922135 \gamma ^2-1417255351884097 \gamma +1844488552375516\right) \mach_\infty^6}{37338428991864000}$ \\\hline
$q_2$ & $\frac{2940401283109290000 \mach_\infty^2 \left[\left(3584503350615 \gamma ^2+3753496299366 \gamma  -1602354228496\right)  \mach_\infty^4-20622030 (102855 \gamma -333287) \mach_\infty^2+72737199334800\right]}{\left[\left(1058753493922135 \gamma ^2-1417255351884097 \gamma +1844488552375516\right) \mach_\infty^6-6561555 (378591917 \gamma -678984653) \mach_\infty^4+14086770937839600 \mach_\infty^2-56007643487796000\right]^2}$ \\\hline
$q_3$ & $\frac{-54894982258481009533016408280000000 \mach_\infty^2 \left[\left(140520538815030 \gamma ^2+13019275653153 \gamma -97154441684485\right) \mach_\infty^4-41244060 (6569315 \gamma +2958729) \mach_\infty^2+678880527124800\right]}{\left[\left(1058753493922135 \gamma ^2-1417255351884097 \gamma+1844488552375516\right) \mach_\infty^6-6561555 (378591917 \gamma -678984653) \mach_\infty^4+14086770937839600 \mach_\infty^2-56007643487796000\right]^3}$ \\\midrule
\end{tabular}

\begin{tabular}{lr}
& $m=4$  \\\midrule
$q_0$ & $\begin{array}{ll} \frac{3}{2}-\frac{83 \mach_\infty^2}{220}+\frac{(378591917 \gamma -678984653) \mach_\infty^4}{5690484800}-\frac{\left(1058753493922135 \gamma ^2-1417255351884097 \gamma +1844488552375516\right) \mach_\infty^6}{37338428991864000} \\ +\frac{\left(14976238371862724157681480000 \gamma ^3-54767356061541106493910208468 \gamma ^2+21478489284009300130011513435 \gamma -44582394863079130273355228987\right) \mach_\infty^8}{ \left(1677449450721507819395719680000\right)} \end{array}$
\\\hline
$q_2$ & $\begin{array}{l} -528396576977274963109651699200000 \mach_\infty^2 \left[\left(113812743442673525009760 \gamma ^3-31571183798299137750086120 \gamma ^2-55929937306277740962895557 \gamma
\right. \right. \nonumber\\ \left. \left.
+20271206182723906597304662\right) \mach_\infty^6-11231387447280 \left(3584503350615 \gamma ^2
+3753496299366 \gamma -1602354228496\right) \mach_\infty^4+231614008879431578400 (102855 \gamma
\right. \nonumber\\  \left.
-333287) \mach_\infty^2-816939667559175886069344000\right] / \left[\left(14976238371862724157681480000 \gamma ^3-54767356061541106493910208468 \gamma ^2
\right. \right. \nonumber\\  \left. \left.
+21478489284009300130011513435 \gamma -44582394863079130273355228987\right) \mach_\infty^8 -44925549789120 \left(1058753493922135 \gamma ^2-1417255351884097 \gamma
\right. \right. \nonumber\\  \left. \left.
+1844488552375516\right) \mach_\infty^6+294781465846549281600 (378591917 \gamma -678984653) \mach_\infty^4-632855929135841586408385152000 \mach_\infty^2
\right. \nonumber\\  \left.
+2516174176082261729093579520000\right]^2 \end{array}$
\\\hline
$q_3$ & $\begin{array}{l} -295452849271218270466030576021899437004399740039626752000000000 \mach_\infty^2\left[\left(1123633481124194863920377280 \gamma ^3-661946767058409457239821580 \gamma ^2
\right. \right. \nonumber\\  \left. \left.
-1086310516712608715079428631 \gamma +804276831488604048889927246\right) \mach_\infty^6-16847081170920 \left(140520538815030 \gamma ^2+13019275653153 \gamma -97154441684485\right)\mach_\infty^4
\right. \nonumber\\  \left.
+694842026638294735200 (6569315 \gamma +2958729) \mach_\infty^2-11437155345828462404970816000\right]/ \left[\left(14976238371862724157681480000 \gamma ^3
\right. \right. \nonumber\\  \left. \left.
-54767356061541106493910208468 \gamma ^2+21478489284009300130011513435 \gamma -44582394863079130273355228987\right) \mach_\infty^8
-44925549789120 \left(1058753493922135 \gamma ^2
\right. \right. \nonumber\\  \left. \left.
-1417255351884097 \gamma+1844488552375516\right) \mach_\infty^6+294781465846549281600 (378591917 \gamma-678984653) \mach_\infty^4-632855929135841586408385152000 \mach_\infty^2
\right. \nonumber\\ \left.
+2516174176082261729093579520000\right]^3 \end{array}$ \\
\bottomrule
\end{tabular}
\end{scriptsize}
\caption{Coefficients of the first four terms in the Taylor expansion of $q\left(u\right)$, for different JRE orders $m$, at general Mach number and $\gamma$.}
\label{table: general hodographic coefficeints}
\end{minipage}}
\end{table}

\section{Details and convergence of the pseudospectral solver}
\label{sec:details and convergence of the pseudospectral solver}

In order to capture the behaviour of the flow far and near the sphere (we review the implementation for the 3D case. The 2D case is similar with a difference only in the differential operators) we map the radial domain from $r\in[1,\infty]$ to $\varrho=r^{-1}\in[0,1]$. The differential operators with respect to $\varrho$ are

\begin{equation}
\label{eq: differential operators for inverse radial coordinate map}
    \grad_\varrho=\left(-\varrho^2\partial_\varrho,\varrho\partial_\theta\right) \quad\mbox{and}\quad \nabla^2_\varrho=\varrho^4\partial_{\varrho\varrho}+\varrho^2\cot{\left(\theta\right)}\partial_\theta+\varrho^2\partial_{\theta\theta}.
\end{equation}

The pseudospectral method requires the expansion of the solution as a series of functions \citep{boyd2001chebyshev} (not required to be orthogonal) and evaluating the equation at collocation points.
To ensure the polar BCs (no polar velocity on the poles \ie Neumann BC for 3D and periodic boundary in 2D) we expand in cosine functions only, furthermore, the back and front reflection symmetry of the sphere limits the cosine functions to only odd cosines (as analytically shown for the JRE in \S\ref{sec:JRE for a hypersphere}), $\cos\left[\left(2n+1\right)\theta\right]$.
In the radial direction we use the Chebyshev polynomials of the first kind (\ref{eq: potential for pseudospectral method}). We take the collocation points to be

\begin{equation}
\label{eq: pseudospectran collocoation points}
    \varrho_i=\cos\left(\frac{\pi i}{2 k_{max}}\right), \; 1 \leq i \leq k_{max}-1; \quad \theta_j=\frac{\pi j}{2\left(n_{max}+2\right)}, \; 1 \leq j \leq n_{max}+1.
\end{equation}

We do not consider the collocation points $\theta\in\{0,1\}$ because the expansion in odd cosines fulfills the BCs identically.
On the points $\varrho\in\{0,1\}$ we solve the BCs at infinity and at the sphere respectively and not the non-linear equation (\ref{eq: normalized flow potential non-linear PDE}).
To solve the non-linear equation we use a simple Newton-Raphson algorithm where at each step we solve a linear set of equations.
In all computation we take the error to be the maximal deviation of the equation at the collocation points and take a tolerance of $10^{-11}$.
Most of the computations converge rapidly and do not require more than seven Newton-Raphson steps.

\begin{figure}
	\centering
	\includegraphics[width=.66\linewidth]{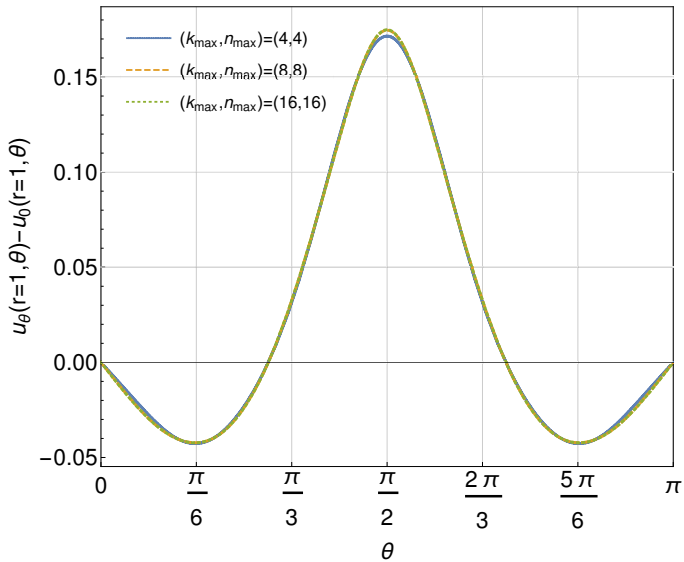}
	\caption{
	Compressibility contribution to the polar velocity on the sphere for different resolutions of the pseudospectral code at $\mach_\infty=0.5619$ and $\gamma=7/5$.}
	\label{fig:sphereconv}
\end{figure}

Figure \ref{fig:sphereconv} shows the compressible contribution to the polar velocity on the sphere for various resolutions of the pseudospectral code.
The code is quite converged, for most angles, except close to the equator and poles.
A resolution of four by four shows the largest relative error (with respect to resolution of 16 by 32), which is smaller than $2\%$ at the equator.
To test the robustness of the computation, we expand the solution with different functions, e.g. odd and even cosines as well as in sine functions, this resulted in the vanishing coefficient of the even cosine and sine functions.
For the radial functions we expand in a linear combination of Chebyshev polynomials which fulfills the BC identically and in Legendre polynomials.
All combinations gave the same results with respect to the tolerance.

We modify the code for the calculation of a compressible flow around spheroids of the form $\left(x^2+y^2\right)/\alpha+z^2/\alpha^2=1$.
We use the prolate spheroidal coordinates, $\mu,\nu$ defined by
\begin{equation}
\label{eq: prolate spheroidal coordinates}
        x=\sqrt{\alpha^2-\alpha}\cosh{(\mu)}\cos{(\nu)}  \quad\mbox{and}\quad
        z=\sqrt{\alpha^2-\alpha}\sinh{(\mu)}\sin{(\nu)} \,.
\end{equation} (we omit the azimutal coordinate $\varphi$ from symmetry considerations).
In this set of coordinates the body is described by the equation $\cosh(\mu_b)=\sqrt{\alpha/(\alpha-1)}$.
We normalize this coordinate and inverse it such that the computational domain is $[0,1]\times[0,\pi/2]$ just as in the case of the sphere.
This changes the differential operators to (after variable change $\mu=\mu_b\varrho^{-1}$):

\begin{eqnarray}
\label{eq: differential operators in prolate spheroidal (and radial inverse map) coordinates}
    \grad_{\varrho}^{(\alpha)}&=&\frac{1}{\sqrt{\left(\alpha^2-\alpha\right)\left[\sinh^2\left(\mu_b\varrho^{-1}\right)+\sin^2\left(\nu\right)\right]}}
    \left(-\varrho^2\mu_b^{-1}\partial_\varrho,\partial_\nu\right) \quad\mbox{and}\quad
    \nonumber\\ \nabla^{2,(\alpha)}_{\varrho}&=&
    \frac{1}{\left(\alpha^2-\alpha\right)\left[\sinh^2\left(\mu_b\varrho^{-1}\right)+\sin^2\left(\nu\right)\right]}\left[\mu_b^{-2}\left(\varrho^4\partial_{\varrho\varrho}+2\varrho^3\partial_\varrho\right)-\varrho^2\mu_b^{-1}\coth\left(\mu_b\varrho^{-1}\right)\partial_\varrho
    \right. \nonumber\\ && \left. +\partial_{\nu\nu}+\cot\left(\nu\right)\partial_\nu\right]
    .
\end{eqnarray}

We take $\phi_0$ to be a function that fulfills both BCs
\begin{equation}
\label{eq: BC fulfilling funtion for prolate spheroidal flow}
    \phi_0^{(\alpha)}=\left(\sqrt{\alpha^2-\alpha}\cosh\left(\mu_b\varrho^{-1}\right)+\sqrt{\alpha}\varrho\mu_b\right)\cos\left(\nu\right).
\end{equation}

 \bibliography{Sphere}
 \bibliographystyle{jfm}
\end{document}